# Nonconvex Nonsmooth Low-Rank Minimization for Generalized Image Compressed Sensing via Group Sparse Representation

Yunyi Li[1], Li Liu[2], Yu Zhao[2], Xiefeng Cheng[1], and Guan Gui[2,*]

[1] College of Electronic and Optical Engineering & College of Microelectronics, Nanjing University of Posts and Telecommunications, Nanjing 210023, China;
[2] College of Telecommunications & Information Engineering, Nanjing University of Posts and Telecommunications, Nanjing 210023, China
*Correspondence Author: guiguan@njupt.edu.cn

**Abstract:** Group sparse representation (GSR) based method has led to great successes in various image recovery tasks, which can be converted into a low-rank matrix minimization problem. As a widely used surrogate function of low-rank, the nuclear norm based convex surrogate usually leads to over-shrinking problem, since the standard soft-thresholding operator shrinks all singular values equally. To improve traditional sparse representation based image compressive sensing (CS) performance, we propose a generalized CS framework based on GSR model, which leads to a nonconvex nonsmooth low-rank minimization problem. The popular $L_2$-norm and M-estimator are employed for standard image CS and robust CS problem to fit the data respectively. For the better approximation of the rank of group-matrix, a family of nuclear norms are employed to address the over-shrinking problem. Moreover, we also propose a flexible and effective iteratively-weighting strategy to control the weighting and contribution of each singular value. Then we develop an iteratively reweighted nuclear norm algorithm for our generalized framework via an alternating direction method of multipliers framework, namely, GSR-AIR. Experimental results demonstrate that our proposed CS framework can achieve favorable reconstruction performance compared with current state-of-the-art methods and the robust CS framework can suppress the outliers effectively.
**Keywords:** group sparse representation; low-rank; nonconvex nonsmooth; nuclear norm; standard CS; robust CS.

## I. Introduction

Image compressive sensing (CS) reconstruction [1] is a classic topic in low-level vision task, which has been widely studied in last decade. It aims at reconstructing a high-quality image **X** from much fewer random measurements **Y**. One of the main technical challenges for CS is how to reduce the measurements, meanwhile, to obtain high-quality images. Typical applications of CS include radar imaging [2], channel estimation in communications systems [3–6], sparse recovery [7] and signal detection [8–12], electrocardiogram signal reconstruction [13], magnetic resonant imaging (MRI) [14–16], and especially in image processing [17–19].

The reconstruction of high-quality images from a small number of measurements is a typical ill-posed inverse problem, it is well-known that the prior knowledge and sparse representation model on image structures play a vital role in the CS reconstruction task. Exploiting more prior knowledge for minimization is often at the core in image CS reconstruction problem. In the past several years, the sparsity-based regularization methods have achieved great success in various CS applications. According to the sparse representation theory, every image can be represented accurately by a few dominant elements in a proper dictionary, which can be learned from natural images or prespecified. Traditional CS recovery approaches use the sparsity of an entire image in a predefined domain and the local structural patterns, such as the famous total-variation (TV) regularization [20]. However, these methods can only exploit a small number of structural features and some important artifacts will appear, e.g., texture and edge information, and thus these methods are not capable of improving reconstruction quality considerably.

As another classic image prior knowledge, recent work has revealed that the non-local similarity of patches can improve the reconstruction quality significantly by exploiting the nonlocal similarity features. For example, the well-known non-local mean filter [21], the promising denoiser BM3D [22], and the standard way for image recovery [23][24]. Recently, a state-of-the art sparse model of non-locally centralized sparse representation (NCSR) [25] is proposed for image restoration, which first obtains an accurate sparse coefficient vector, and then centralizes the sparse coefficients to enhance the sparsity and improve the performance. Recent advances suggest that the group sparse representation (GSR) based approaches can often lead to great improvements by removing the artifacts and preserving the details. The group-matrix is constructed using similar patches and hence owns the low-rank property, then the low-rank is a useful image prior for image restoration [26].

For dealing with the image CS reconstruction problem, another important issue is how to regularize the sparsity. Conventional methods use the $L_1$-norm as the surrogate of the $L_0$-norm, and the resulting convex optimization problem can be easily solved. However, the achieved solution by $L_1$-norm regularization is usually suboptimal to the $L_0$-norm based minimization because of this loose appropriation. Hence, to appropriate the $L_0$-norm by nonconvex function will achieve a more accurate solution [27][28]. Typical nonconvex surrogate function including $L_p$-norm [29–31], Smoothly Clipped Absolute Deviation (SCAD) [32], Logarithm [33], and Minimax Concave Penalty (MCP) [34], etc.. Recently, the low-rank based regularization approaches have shown its great potentials in image processing [35][36], especially in CS image recovery [37][38]. It is the fact that the adjacent patches in an image have the similar structures, if several similar patches are constructed as a group-matrix, then the matrix shows the low-rank property. Hence, the image CS problem can convert into a low-rank matrix approximation problem. Since the low-rank minimization problem is a NP-hard problem, it is usually relaxed as the nuclear norm minimization (NNM) problem [26].

Standard CS framework reconstructs image from CS measurements under Gaussian noise, however, in real world applications, CS observations are usually corrupted by impulsive noise. It is not efficient for standard CS to suppress the outliers because the $L_2$-norm based fidelity term is very sensitive to the outliers. To recovery signal or image from CS measurements under impulsive noise, many efficient algorithms have been proposed, such as the $L_1$-norm YALL1 [39], LA-Lq [40], the M-estimator based robust CS recovery [41]. However, these robust CS methods are all based on traditional sparse signal recovery framework, and ignore important structure information of images. Recently, a novel robust CS method for image is proposed, which has shown promising performances than traditional robust sparse recovery methods for image CS problems. It can exploit the latent structure and sparse prior for minimization, such the non-local similarity of nature image [42].

In this paper, inspired by the successes of nonconvex regularization approaches and the promising GSR model, we first propose a GSR based denoising model, and then our denoising model can be utilized for image CS reconstruction framework. To enhance the low-rank reconstruction performance, we extend a family of typical nonconvex surrogate penalties of $L_0$-norm on singular values of the group matrix problem, which leads a generalized low-rank minimization problem. The main contributions of our work can be concluded as follows. We first build the connection between GSR based denoising model and the low-rank minimization problem, and then propose a generalized GSR based CS framework, in which, the local sparsity and nonlocal similarity of image can be unified simultaneously by this framework. For standard CS problem, we employ the popular $L_2$-norm as the fidelity term, while a M-estimator is utilized to suppress the outliers caused by impulsive noise for robust CS problem. To deal with the resulting nonconvex nonsmooth optimization problem, we develop an iteratively reweighted nuclear norm algorithm based on alternative direction method of multipliers, namely GSR-AIR. More importantly, we also propose an iteratively-weighting strategy to control the weighting and contribution on each singular value. At last, we evaluate our proposed nonconvex framework by using several well-known nonconvex penalty functions of ETP, logarithm, $L_p$ and MCP on the classic image CS reconstruction problem.

The rest of the paper is organized as follows. In the second section, we will first introduce the GSR theory, and then build a connection between GSR based denoising model and low-rank minimization problem. In the section III, we propose an image CS framework via GSR based denoising model, and then a family of nonconvex and nonsmooth surrogate functions are adopted to enhance the low-rank matrix recovery, we will detail our proposed GSR-AIR algorithm. Section IV provides simulation results compared with current state-of-the-art methods to demonstrate the effectiveness and priority of our proposed framework. Finally, a brief summary will be concluded in section V.

## II. Group Sparse Representation based Denoising Model via Low-Rank Minimization

Traditional patch-based sparse representation modeling is inaccurate, because each patch is considered separately and the relationship among patches is ignored. The new GSR model can represent the image in the domain of group, which not only enhances intrinsic local sparsity, but also enhances the nonlocal similarity simultaneously [26].

### 2.1. Group Sparse Representation

For image $\mathbf{X} \in \mathbb{R}^{\sqrt{N} \times \sqrt{N}}$, which can be divided into $n$ overlapped patches $\mathbf{X}_k, k = 1, 2, \cdots, n$ with the size of $\sqrt{\mathcal{B}_s} \times \sqrt{\mathcal{B}_s}, \mathcal{B}_s < N$. Given a searching window with the size of $L \times L$ for each patch $\mathbf{X}_k$, then we can search $c$ best matched patches by using the well-known Euclidean distance as the similarity criterion, and the set of these best similar patches denotes $S_{\mathbf{x}_k}$. Next, these $c$ best matched patches are stacked into a matrix with the size of $\mathcal{B}_s \times c$, denoted by $\mathbf{X}_{G_k} = [\mathbf{X}_{G_k,1}, \mathbf{X}_{G_k,2}, \cdots, \mathbf{X}_{G_k,c}] \in \mathbb{R}^{\mathcal{B}_s \times c}$, where each patch can be vectorized as $\mathbf{X}_{G_k,i} \in \mathbb{R}^{\mathcal{B}_s \times 1}, i = 1, 2, \cdots, c$ as the columns. Such matrix $\mathbf{X}_{G_k}$ with $c$ patches containing similar structures is named as group, we define the construction process of group as $\mathbf{X}_{G_k} = G_k(\mathbf{X})$, where the operator $G_k(\cdot)$ denotes the group construction operator from $\mathbf{X}$. A simple illustration of the group matrix construction process is presented is the Fig. 1. Different from the traditional patch-based sparse representation model, the GSR model can exploit the nonlocal self-similarity of image and enhance the local sparsity by using the group as basic unit for sparse representation [26].

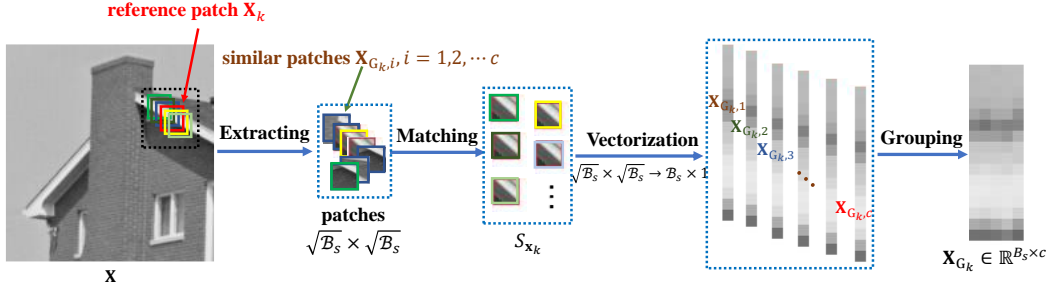

**Fig. 1**. A simple illustration of the group matrix construction process.

According to the sparse representation theory, the reconstructed group can be represented sparsely by

$$\mathbf{X}_{G_k} = \mathbf{D}_{G_k}\boldsymbol{\alpha}_{G_k} = \sum_{i=1}^{m}\alpha_{G_k,i}\boldsymbol{d}_{G_k,i} \quad (1)$$

where $\mathbf{D}_{G_k} = [\boldsymbol{d}_{G_k,1}, \boldsymbol{d}_{G_k,2}, \cdots, \boldsymbol{d}_{G_k,m}] \in \mathbb{R}^{(B_s \times c) \times m}$ denotes a 3D dictionary, and each atom $\boldsymbol{d}_{G_k,i} \in \mathbb{R}^{B_s \times c}, i = 1,2,\cdots,m$ is a matrix with the same size of each group $\mathbf{X}_{G_k}$, $\boldsymbol{\alpha}_{G_k} = [\alpha_{G_k,1}, \alpha_{G_k,2}, \cdots, \alpha_{G_k,m}] \in \mathbb{R}^{m \times 1}$ is a vector. When all the dictionary $\{\mathbf{D}_{G_k}\}, k = 1,2,\cdots,n$ and the sparse codes $\{\boldsymbol{\alpha}_{G_k}\}, k = 1,2,\cdots,n$ are known, the whole image $\mathbf{X}$ can be represented by

$$\mathbf{X} = \mathbf{D}_G \circ \boldsymbol{\alpha}_G \quad (2)$$

where the dictionary $\mathbf{D}_G$ and sparse code $\boldsymbol{\alpha}_G$ denote the concatenation of all $\mathbf{D}_{G_k}$ and $\boldsymbol{\alpha}_{G_k}$ respectively.

**2.2. Adaptive dictionary learning**

To obtain an adaptive dictionary $\mathbf{D}_{G_k}$ for each group $\mathbf{X}_{G_k}$, in this paper, we adopt a self-adaptive dictionary learning scheme [26][43] for each group, we can learn the adaptive dictionary $\mathbf{D}_{G_k}$ from $\mathbf{X}_{G_k}$ directly. We first employ the singular value decomposition (SVD) of $\mathbf{X}_{G_k}$ by

$$\mathbf{X}_{G_k} = \mathbf{U}_{\mathbf{X}_{G_k}} \boldsymbol{\Sigma}_{\mathbf{X}_{G_k}} \mathbf{V}_{\mathbf{X}_{G_k}}^T = \sum_{i=1}^{m} \sigma_{\mathbf{X}_{G_k},i} \boldsymbol{u}_{\mathbf{X}_{G_k},i} \boldsymbol{v}_{\mathbf{X}_{G_k},i}^T \quad (3)$$

where $m = min(B_s, c)$, $\mathbf{U}_{\mathbf{X}_{G_k}} = [\boldsymbol{u}_{\mathbf{X}_{G_k},1}, \boldsymbol{u}_{\mathbf{X}_{G_k},2}, \cdots, \boldsymbol{u}_{\mathbf{X}_{G_k},m}] \in \mathbb{R}^{B_s \times m}$, $\boldsymbol{\Sigma}_{\mathbf{X}_{G_k}} = diag([\sigma_{\mathbf{X}_{G_k},1}; \sigma_{\mathbf{X}_{G_k},2}; \cdots; \sigma_{\mathbf{X}_{G_k},m}]) \in \mathbb{R}^{m \times m}$ and $\mathbf{V}_{\mathbf{X}_{G_k}} = [\boldsymbol{v}_{\mathbf{X}_{G_k},1}, \boldsymbol{v}_{\mathbf{X}_{G_k},2}, \cdots, \boldsymbol{v}_{\mathbf{X}_{G_k},m}] \in \mathbb{R}^{m \times c}$. Then, the $i$-th atom $\boldsymbol{d}_{G_k,i}$, $i = 1,2,\cdots,m$ of the dictionary $\mathbf{D}_{G_k}$ can be obtained by

$$\boldsymbol{d}_{G_k,i} = \boldsymbol{u}_{\mathbf{X}_{G_k},i} \boldsymbol{v}_{\mathbf{X}_{G_k},i}^T \quad (4)$$

Finally, we can achieve the self-adaptive dictionary for each group by

$$\mathbf{D}_{G_k} = [\boldsymbol{d}_{G_k,1}, \boldsymbol{d}_{G_k,2}, \cdots, \boldsymbol{d}_{G_k,m}]. \quad (5)$$

**2.3. GSR based denoising model**

After achieving the adaptive dictionary $\mathbf{D}_{G_k}$ for each group, then the GSR based denoising model from the degraded observation $\mathbf{Y}_{G_k} \in \mathbb{R}^{B_s \times c}$ can be formulated as

$$\widehat{\boldsymbol{\alpha}}_{G_k} = \arg\min_{\boldsymbol{\alpha}_{G_k}} \frac{1}{2} \|\mathbf{Y}_{G_k} - \mathbf{D}_{G_k}\boldsymbol{\alpha}_{G_k}\|_F^2 + \lambda \|\boldsymbol{\alpha}_{G_k}\|_0 \quad (6)$$

where $\boldsymbol{\alpha}_{G_k}$ denotes the sparse coefficient vector over the dictionary $\mathbf{D}_{G_k}$, and $\lambda$ denotes the regularization parameter.

According to the definition (1) and (3), the number of nonzero elements of vector $\boldsymbol{\alpha}_{G_k}$ is equal to the number of numbers of nonzero singular values of $\mathbf{X}_{G_k}$. Then we have the following theory

**Theorem 2.1.** *For image group $\mathbf{X}_{G_k}$ and its singular value decomposition (SVD) $\mathbf{X}_{G_k} = \sum_{i=1}^{m} \sigma_{\mathbf{X}_{G_k},i} \boldsymbol{u}_{\mathbf{X}_{G_k},i} \boldsymbol{v}_{\mathbf{X}_{G_k},i}^{\mathrm{T}}$, if $\mathbf{X}_{G_k} = \mathbf{D}_{G_k} \boldsymbol{\alpha}_{G_k}$, then the number of nonzero elements of $\boldsymbol{\alpha}_{G_k}$ is equal to the number of nonzero value of singular values $\sigma_{\mathbf{X}_{G_k},i}$, we have the following relationship*

$$\|\boldsymbol{\alpha}_{G_k}\|_0 = \|\boldsymbol{\sigma}_{\mathbf{X}_{G_k}}\|_0 = \mathrm{rank}(\mathbf{X}_{G_k}) \tag{7}$$

where the sparse vector $\boldsymbol{\alpha}_{G_k} = [\alpha_{G_k,1}, \alpha_{G_k,2}, \cdots, \alpha_{G_k,m}] \in \mathbb{R}^{m \times 1}$, the singular value vector $\boldsymbol{\sigma}_{\mathbf{X}_{G_k}} = [\sigma_{\mathbf{X}_{G_k},1}, \sigma_{\mathbf{X}_{G_k},2}, \cdots, \sigma_{\mathbf{X}_{G_k},m}] \in \mathbb{R}^{m \times 1}$, $\mathbf{D}_{G_k} = [\boldsymbol{d}_{G_k,1}, \cdots, \boldsymbol{d}_{G_k,m}] \in \mathbb{R}^{(B_s \times c) \times m}$, and $\mathrm{rank}(\mathbf{X}_{G_k})$ denotes the singular value number of the matrix $\mathbf{X}_{G_k}$.

Accordingly, by substituting $\mathbf{X}_{G_k} = \mathbf{D}_{G_k} \boldsymbol{\alpha}_{G_k}$ in (6), the denoising problem (6) has the following equivalent low-rank minimization problem

$$\widehat{\mathbf{X}}_{G_k} = \arg\min_{\mathbf{X}_{G_k}} \frac{1}{2} \|\mathbf{Y}_{G_k} - \mathbf{X}_{G_k}\|_F^2 + \lambda Rank(\mathbf{X}_{G_k}). \tag{8}$$

where $\mathbf{X}_{G_k}$ denotes the constructed image group with low-rank property. Then we can convert the sparsity-inducing optimization problem (6) into the low-rank minimization problem (8).

### III. Image Compressed Sensing via GSR based denoising model

In general, the image CS observation model can be expressed as

$$\mathbf{Y} = \mathbf{HX} + \mathbf{n} \tag{9}$$

where $\mathbf{H} \in \mathbb{R}^{M \times N}, (M \ll N)$ denotes the random measurement matrix, $\mathbf{X} \in \mathbb{R}^N$ (also $\mathbf{X} \in \mathbb{R}^{\sqrt{N} \times \sqrt{N}}$) denotes the desired image, and $\mathbf{Y} \in \mathbb{R}^M$ denotes the measurements which can be corrupted by the noise $\mathbf{n} \in \mathbb{R}^M$. When the original image $\mathbf{X}$ can be sparsely represented by a given dictionary $\mathbf{D}$, denotes as $\mathbf{X} = \mathbf{D}\boldsymbol{\alpha}$, then the CS reconstruction problem can be resolved by the following regularization method

$$\widehat{\boldsymbol{\alpha}} = \arg\min_{\boldsymbol{\alpha}} f(\mathbf{Y} - \mathbf{HD}\boldsymbol{\alpha}) + \lambda \|\boldsymbol{\alpha}\|_0 \tag{10}$$

where $f(\mathbf{Y} - \mathbf{HD}\boldsymbol{\alpha})$ denotes the data fidelity term, e.g., the $L_2$-norm for Gaussian noise, and the M-estimator [44] to fit the data under impulsive noise, the $L_0$-norm denotes regularization term, which measures the sparsity degree of image, and can provide the necessary prior knowledge for minimization, the regularization parameter $\lambda$ controls the tradeoff between the fidelity term and the regularization term. With the achieved $\widehat{\boldsymbol{\alpha}}$, we can reconstruct the latent image by

$$\widehat{\mathbf{X}} = \mathbf{D}\widehat{\boldsymbol{\alpha}} \tag{11}$$

Accordingly, after stacking the related similar patches to generate the group $\mathbf{X}_{G_k} \in \mathbb{R}^{B_s \times c}, k = 1, 2, \cdots, n$, then the group based measurement model of whole image $\mathbf{X}$ is formulated as

$$\mathbf{Y} = \mathbf{HD}_G \circ \boldsymbol{\alpha}_G + \mathbf{n} \tag{12}$$

Then, the GSR model based image CS problem can be written as,

$$\hat{\boldsymbol{\alpha}}_G = \arg\min_{\boldsymbol{\alpha}_G} f(\mathbf{Y} - \mathbf{H}\mathbf{D}_G \circ \boldsymbol{\alpha}_G) + \lambda \|\boldsymbol{\alpha}_G\|_0 \qquad (13)$$

where $\|\boldsymbol{\alpha}_G\|_0 = \sum_{k=1}^{n} \|\boldsymbol{\alpha}_{G_k}\|_0$. According to the relationship in (7), let $\mathbf{X} = \mathbf{D}_G \circ \boldsymbol{\alpha}_G$, then the optimization problem (13) can be turned into the following low-rank minimization problem

$$\hat{\mathbf{X}} = \arg\min_{\mathbf{X}} f(\mathbf{Y} - \mathbf{H}\mathbf{X}) + \lambda \sum_{k=1}^{n} Rank(\mathbf{X}_{G_k}) \qquad (14)$$

where $\sum_{k=1}^{n} Rank(\mathbf{X}_{G_k})$ denotes the sum of all low-rank matrixes for each $\mathbf{X}$.

### 3.1. Nonconvex nonsmooth low-rank minimization for image CS reconstruction

It is often a challenge problem to solve the above low rank optimization problem (14) and the rank function is usually relaxed as the convex nuclear norm, after replacing by the popular convex nuclear norm, the NNM based optimization problem can be expressed as

$$\hat{\mathbf{X}} = \arg\min_{\mathbf{X}} f(\mathbf{Y} - \mathbf{H}\mathbf{X}) + \lambda \sum_{k=1}^{n} \|\mathbf{X}_{G_k}\|_* \qquad (15)$$

where $\|\mathbf{X}_{G_k}\|_* = \sum_i |\sigma_i(\mathbf{X}_{G_k})|$ denotes the nuclear norm, and $\sigma_i(\mathbf{X}_{G_k}), i = 1,2,\cdots d, d = min(B_s, c)$ are the singular values of matrix $\mathbf{X}_{G_k}$. Although the above model (15) can incorporate the low rank prior knowledge, the NNM usually treats different rank components (singular values) equally and simultaneously, hence it cannot achieve the approximation of the low-rank accurately. Recently, the nonconvex penalized regularization methods have shown great potential to improve the sparse recovery performance, typical nonconvex surrogate functions include the $L_p$ function [30], Smoothly Clipped Absolute Deviation (SCAD) [32], Logarithm function [33], and Minimax Concave Penalty (MCP) [34], etc.. Although the nonconvex strategy can improve the NNM effectively, such as a recent proposed work based on nonconvex $L_p$ nuclear norm for CS problem [45], it still has some problems. According to the theory of low rank minimization, the rank of a certain matrix only corresponds to the larger nonzero singular values, what's more, larger singular values often contain more information of matrix. To approximate the rank of the group-matrix more accurately, hence, the larger singular values should be shrunk less, and the smaller ones should be shrunk more. In this paper, we extend a class of nonconvex nonsmooth functions to regularize the sparsity of singular values, then our proposed nonconvex nonsmooth weighted framework can be expressed as

$$\hat{\mathbf{X}} = \arg\min_{\mathbf{X}} f(\mathbf{Y} - \mathbf{H}\mathbf{X}) + \lambda \sum_{k=1}^{n} \Re(\mathbf{X}_{G_k}) \qquad (16)$$

where $\Re(\mathbf{X}_{G_k}) = \sum_{i=1}^{r} \rho(\sigma_i(\mathbf{X}_{G_k}))$, $\rho(\cdot): \mathbb{R}^+ \to \mathbb{R}^+$ denotes the nonconvex nonsmooth penalty functions, and $\sigma_i(\mathbf{X}_{G_k}), i = 1,2,\cdots,r = min(B_s, c)$ denote singular values with $\sigma_1(\mathbf{X}_{G_k}) \geq \sigma_2(\mathbf{X}_{G_k}) \geq \cdots \geq \sigma_r(\mathbf{X}_{G_k})$. For our proposed framework, some classic nonconvex surrogate functions of $\|\boldsymbol{\theta}\|_0$ are detailed as the **Definition 1** and **Fig**. 2.

It should be noted that our work is not a simple extend version compared with existing work [45][46]. Compared with [45] and [46], we not only extend the nonconvex $L_p$-norm into a family of nonconvex surrogate functions to regularize the singular values of group matrix, such as ETP, Logarithm, MCP, SCAD, etc. More significantly, we propose our CS framework based on our proposed denoising model, where we provide theorical connection between GSR model and our denoising model. What's more, we also address the robust image CS reconstruction problem under non-Gaussian noise environments.

**Definition 1** *Popular and typical nonconvex surrogate functions* $\rho(\theta)$ *and their super-gradients* $\partial \rho(\theta)$

| Penalty | Formula $\rho(\theta)$ | Super-gradient $\partial\rho(\theta)$ |
|---|---|---|
| $L_p$ [30] | $L_p(\theta) = \lambda\theta^p$ | $\begin{cases} \infty, & if\ \theta = 0 \\ \lambda p \theta^{p-1}, & if\ \theta > 0 \end{cases}$ |
| SCAD [32] | $\begin{cases} \lambda\theta, & if\ \theta \leq \lambda \\ \dfrac{-\theta^2 + 2\gamma\lambda\theta - \lambda^2}{2(\gamma - 1)}, & if\ \lambda < \theta \leq \gamma\lambda \\ \dfrac{\lambda^2(\gamma + 1)}{2}, & if\ \theta > \gamma\lambda \end{cases}$ | $\begin{cases} \lambda, & if\ \theta = 0 \\ \dfrac{\gamma\lambda - \theta}{\gamma - 1}, & if\ \lambda < \theta \leq \gamma\lambda \\ 0, & if\ \theta > \gamma\lambda \end{cases}$ |
| Logarithm [33] | $\dfrac{\lambda}{\log(\gamma + 1)}\log(\gamma\theta + 1)$ | $\dfrac{\gamma\lambda}{(\gamma\theta + 1)\log(\gamma + 1)}$ |
| MCP [34] | $\begin{cases} \lambda\theta - \dfrac{\theta^2}{2\gamma}, & if\ \theta < \gamma\lambda \\ \dfrac{\gamma\lambda^2}{2}, & if\ \theta > \gamma\lambda \end{cases}$ | $\begin{cases} \lambda - \dfrac{\theta}{\gamma}, & if\ \theta < \gamma\lambda \\ 0, & if\ \theta \geq \gamma\lambda \end{cases}$ |
| ETP [47] | $\dfrac{\lambda}{1 - e^{-\gamma}}(1 - e^{-\gamma\theta})$ | $\dfrac{\lambda\gamma}{1 - e^{-\gamma}}e^{-\gamma\theta}$ |
| Capped $L_1$ [48] | $\begin{cases} \lambda\theta, & if\ \theta < \gamma \\ \lambda\gamma, & if\ \theta \geq \gamma\lambda \end{cases}$ | $\begin{cases} \lambda, & if\ \theta < \gamma \\ [0, \lambda], & if\ \theta = \gamma \\ 0, & if\ \theta \geq \gamma\lambda \end{cases}$ |
| Geman [49] | $\dfrac{\lambda\theta}{\theta + \gamma}$ | $\dfrac{\lambda\gamma}{(\theta + \gamma)^2}$ |
| Laplace [50] | $\lambda\left(1 - e^{-\frac{\theta}{\gamma}}\right)$ | $\dfrac{\lambda}{\gamma}e^{-\frac{\theta}{\gamma}}$ |

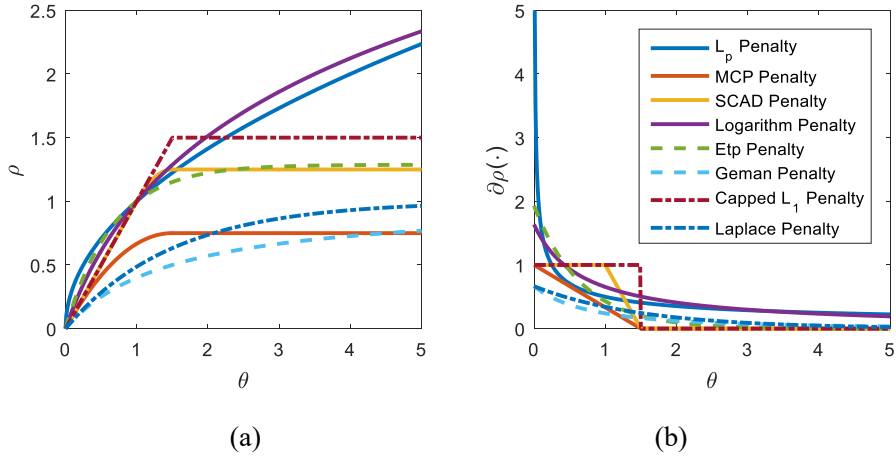

**Fig. 2**. Illustrations of eight typical nonconvex penalties $\varphi(x)$, and their corresponding super-gradients $\partial[\varphi(x)]$. For all penalties, $\lambda = 1$ and $\gamma = 1.5$. (a) Eight typical nonconvex penalties $\varphi(x)$; (b) Their corresponding super-gradients $\partial[\varphi(x)]$.

### 3.2. Proposed GSR-AIR algorithm

The alternative direction method of multipliers (ADMM) framework is an efficient and effective approach for large-scale optimization problem, which can split the constrained minimization problem into several constrained sub-problems [51]. To solve the nonconvex nonsmooth optimization problem (16), we introduce a GSR-AIR algorithm which consists of three stages: Firstly, the ADMM framework is applied to our non-convex minimization problem (16) to split it into three subproblems which is more effective and efficient than solving it directedly. Secondly, we employ an iteratively-reweighted nuclear norm (IRNN) algorithm for the $\mathbf{X}_{G_k}$-subproblem by observing that all gradients of nonconvex surrogate function are nonnegative and monotonically increasing in $[0, \infty)$. Finally, we propose a flexible and effective weighting strategy to address the over-shrinking problem.

In the first stage, we first consider the optimization model (16), by introducing the auxiliary variable $\mathbf{Z}$ with the constraint $\mathbf{X} = \mathbf{Z}$, i.g.,

$$\widehat{\mathbf{X}} = \arg\min_{\mathbf{X}} f(\mathbf{Y} - \mathbf{HX}) + \lambda \sum_{k=1}^{n} \Re(\mathbf{Z}_{G_k}), \quad \text{s.t.} \quad \mathbf{X} = \mathbf{Z} \tag{17}$$

then we have the following three iterative steps

$$\mathbf{X}^{(t+1)} = \arg\min_{\mathbf{X}} f(\mathbf{Y} - \mathbf{HX}) + \frac{\mu}{2} \left\| \mathbf{X} - \mathbf{Z}^{(t)} - \mathbf{W}^{(t)} \right\|_2^2, \tag{18}$$

$$\mathbf{Z}^{(t+1)} = \arg\min_{\mathbf{Z}} \frac{\mu}{2} \left\| \mathbf{X}^{(t+1)} - \mathbf{Z} - \mathbf{W}^{(t)} \right\|_2^2 + \lambda \sum_{k=1}^{n} \Re(\mathbf{Z}_{G_k}), \tag{19}$$

$$\mathbf{W}^{(t+1)} = \mathbf{W}^{(t)} - \left( \mathbf{X}^{(t+1)} - \mathbf{Z}^{(t+1)} \right). \tag{20}$$

Then the optimization problem (16) can be split into three sub-problems (18) to (20).

**(A). X-subproblem**

The solution of optimization problem (18) is not definite because the formulation of $f(\mathbf{Y} - \mathbf{HX})$ is not decided. This paper addresses the standard image CS problem under gaussian environment and the robust image CS problem impulsive noise environment. For noiseless and gaussian noise environment, we employ $f(\mathbf{Y} - \mathbf{HX}) = \frac{1}{2} \|\mathbf{Y} - \mathbf{HX}\|_2^2$ as the fidelity term to fit the data, thus the problem (18) is written as

$$\mathbf{X}^{(t+1)} = \arg\min_{\mathbf{X}} \frac{1}{2} \|\mathbf{Y} - \mathbf{HX}\|_2^2 + \frac{\mu}{2} \left\| \mathbf{X} - \mathbf{Z}^{(t)} - \mathbf{W}^{(t)} \right\|_2^2 \tag{21}$$

The **X**-subproblem of (21) is a strictly convex minimization problem, which has a closed-form solution expressed as

$$\mathbf{X} = (\mathbf{H}^T \mathbf{H} + \mu \mathbf{I})^{-1} (\mathbf{H}^T \mathbf{Y} + \mu (\mathbf{X} + \mathbf{W})) \tag{22}$$

where $\mathbf{I}$ denotes the identity matrix. It is often inefficient to achieve the solution by (22) directly for CS reconstruction problem, since without specific structure of observation matrix $\mathbf{H}$. To avoid the computing of matrix inverse, here, we employ the gradient descent method to solve the sub-problem of (21) by [26]

$$\widetilde{\mathbf{X}} = \mathbf{X} - \eta \boldsymbol{d} \tag{23}$$

where the parameter $\eta$ is the optimal step, $\boldsymbol{d}$ denotes the gradient direction of $\frac{1}{2} \|\mathbf{Y} - \mathbf{HX}\|_2^2 + \frac{\mu}{2} \|\mathbf{X} - \mathbf{Z} - \mathbf{W}\|_2^2$, then we have

$$d = H^T HX - H^T Y + \mu(X - Z - W). \tag{24}$$

When the measurement is corrupted by impulsive noise, e.g., gaussian mixture noise [41], we utilize the M-estimator [44] to fit the data, thus

$$X^{(t+1)} = \arg\min_{X} \frac{1}{2}\psi(Y - HX) + \frac{\mu}{2}\|X - Z^{(t)} - W^{(t)}\|_2^2 \tag{25}$$

where $\psi(Y - HX) = \sum_{i=1}^{M} \varphi((Y - HX)_i)$ with $\varphi(\cdot) = 1 - exp\left(-\frac{(\cdot)^2}{\sigma^2}\right)$, the operator $(\cdot)_i$ denotes the $i$-th element of vector.

**Theorem 3.1** [52]. *Considering the following optimization problem*

$$\min_{X} \varphi(X) + \mathcal{R}(X) \tag{26}$$

*where $\varphi(X) = \sum_i \varphi(X_i)$ denotes the potential loss function in half-quadratic, $\mathcal{R}(X)$ is a convex penalty term, the optimization problem can be reformulated as for a fixed*

$$\min_{X}\{Q(X, q) + \sum_i \emptyset(q_i)\} + \mathcal{R}(X) \tag{27}$$

*where $\emptyset(\cdot)$ denotes the conjugate function of $\varphi(\cdot)$.*

According to the **Theorem 3.1**, the subproblem can be reformulated as

$$(X^{(t+1)}, q^{(t+1)}) = \arg\min_{X,q} \left\{ Q(Y - HX^{(t)}, q^{(t)}) + \sum_{i=1}^{M} \emptyset(q_i^{(t)}) + \mu\|X - Z^{(t)} - W^{(t)}\|_2^2 \right\} \tag{28}$$

the above joint optimization problem can be resolved by optimizing the following two problems iteratively, as

$$q^{(t+1)} = \arg\min_{q} Q(Y - HX^{(t)}, q) + \sum_{i=1}^{M} \emptyset(q_i) \tag{29}$$

and

$$X^{(t+1)} = \arg\min_{X} Q(Y - HX^{(t)}, q^{(t)}) + \mu\|X - Z^{(t)} - W^{(t)}\|_2^2. \tag{30}$$

For the sub-problem of $q$, we have

$$q^{(t+1)} = \exp\left(-\frac{(Y - HX^{(t)})_i^2}{\sigma^2}\right) \tag{31}$$

For the sub-problem of $X$,

$$X^{(t+1)} = \arg\min_{X} \frac{1}{2} Q(Y - HX, q^{(t+1)}) + \frac{\mu}{2}\|X - Z^{(t)} - W^{(t)}\|_2^2 \tag{32}$$

After replacing $Q(Y - HX, q^{(t+1)})$ with $\sum_{i=1}^{m} q_i^{(t+1)}(Y - HX)_i^2$, the following optimization problem is obtained,

$$X^{(t+1)} = \arg\min_{X} \frac{1}{2} \left\| \sqrt{Q^{(t+1)}}(Y - HX) \right\|_2^2 + \frac{\mu}{2}\|X - Z^{(t)} - W^{(t)}\|_2^2 \tag{33}$$

where $Q$ is a diagonal matrix with its diagonal elements $Q_{ii} = q_i$. To solve the above optimization problem more efficiently, we also implement it using the gradient descent method, that is

$$\widetilde{X} = X - \eta d \tag{34}$$

with

$$d = H^T Q^{(t+1)} HX - H^T Q^{(t+1)} Y + \mu(X - Z - W). \tag{35}$$

**(B). Z-subproblem**

In the second stage, we will solve the **Z**-subproblem by the famous IRNN algorithm [53]. After achieving **X**, the **Z**-subproblem can be expressed as

$$\mathbf{Z}^{(t+1)} = \arg\min_{\mathbf{Z}} \frac{\mu}{2}\|\mathbf{R}^{(t+1)} - \mathbf{Z}\|_2^2 + \lambda \sum_{k=1}^{n} \Re(\mathbf{Z}_{G_k}) \tag{36}$$

where $\mathbf{R}^{(t+1)} = \mathbf{X}^{(t+1)} - \mathbf{W}^{(t)}$ and $\Re(\mathbf{Z}_{G_k}) = \sum_{i=1}^{r} \rho\left(\sigma_i(\mathbf{Z}_{G_k})\right)$. The problem of (36) is a typical denoising problem, where **R** denotes the noisy observation of **X** [38]. However, it is difficult to solve it because of the complicated structure of regularizer. By grouping the similar patches of **R** to generate the $\mathbf{R}_{G_k} \in \mathbb{R}^{\mathcal{B}_s \times c}$, according to the relationship between $\sum_{k=1}^{n}\|\mathbf{R}_{G_k} - \mathbf{Z}_{G_k}\|_F^2$ and $\|\mathbf{R} - \mathbf{Z}\|_2^2$, then we have [54]

$$\|\mathbf{R} - \mathbf{Z}\|_2^2 = \frac{N}{K}\sum_{k=1}^{n}\|\mathbf{R}_{G_k} - \mathbf{Z}_{G_k}\|_F^2 \tag{37}$$

where $K = n \times c \times B_s$. Then the problem of (36) can be transformed into the following $n$ subproblems

$$\mathbf{Z}^{(t+1)} = \arg\min_{\mathbf{Z}_{G_k}} \frac{1}{2}\sum_{k=1}^{n}\left\{\|\mathbf{R}_{G_k} - \mathbf{Z}_{G_k}\|_F^2 + \tau \sum_{k=1}^{n} \Re(\mathbf{Z}_{G_k})\right\} \tag{38}$$

where $\tau = \frac{\lambda K}{\mu N}$. For each group $\mathbf{Z}_{G_k}, k = 1,2,\cdots,n$.

### 3.3 GSR-IRNN algorithm for denoising model

This subsection will develop a GSR-IRNN algorithm for subproblem (38). For each group $\mathbf{Z}_{G_k}, k = 1,2,\cdots,n$, the GSR model based denoising problem for $\mathbf{Z}_{G_k}$ is formulated as

$$\mathbf{Z}_{G_k}^{(t+1)} = \arg\min_{\mathbf{Z}_{G_k}} \frac{1}{2}\|\mathbf{R}_{G_k} - \mathbf{Z}_{G_k}\|_F^2 + \tau \Re(\mathbf{Z}_{G_k}). \tag{39}$$

where $\Re(\mathbf{Z}_{G_k}) = \sum_{i=1}^{r} \rho\left(\sigma_i(\mathbf{Z}_{G_k})\right)$. According to the **Definition 1**, we can observe that all the nonconvex functions contain common properties: concave and monotonically increasing on $[0,\infty)$. We first give the definition of super-gradient for all the nonconvex functions defined in **Definition 1**. We can easily observe that their super-gradients are nonnegative and monotonically decreasing, thus we can propose a general solver for the problem (39).

**Theorem 3.1.** [55] *Let* $g: \mathbb{R}^n \to \mathbb{R}$ *be concave, if for every* $\mathbf{R} \in \mathbb{R}^n$, *a vector* $\mathbf{v}$ *is a super-gradient of* $g(\mathbf{Z})$ *at the point* $\mathbf{Z} \in \mathbb{R}^n$, *then*

$$g(\mathbf{R}) \leq g(\mathbf{Z}) + \langle \mathbf{v}, \mathbf{R} - \mathbf{Z} \rangle. \tag{40}$$

In this paper, since the function of $g(\cdot)$ is concave on $[0,\infty)$, according to the **Definition 2** of the super-gradient, and the **Theorem 2**, we can have

$$\rho\left(\sigma_i(\mathbf{Z}_{G_k})\right) \leq \rho\left(\sigma_i(\mathbf{Z}_{G_k}^t)\right) + \widetilde{\omega}_{k,i}^t \left(\sigma_i(\mathbf{Z}_{G_k}) - \sigma_i(\mathbf{Z}_{G_k}^t)\right) \tag{41}$$

where $\widetilde{\omega}_{k,i}^t \in \partial\left[\rho\left(\sigma_i(\mathbf{Z}_{G_k}^t)\right)\right]$, termed as the reweight here. Since $\sigma_1(\mathbf{Z}_{G_k}^t) \geq \sigma_2(\mathbf{Z}_{G_k}^t) \geq \cdots \geq \sigma_m(\mathbf{Z}_{G_k}^t) \geq 0$, then according to the antimonotone property of super-gradient defined in **Table 1**, we have

$$0 \leq \widetilde{\omega}_{k,1}^t \leq \widetilde{\omega}_{k,2}^t \leq \cdots \leq \widetilde{\omega}_{k,m}^t \tag{42}$$

where $\mathbf{Z}_{G_k}^t$ denotes the $t$-th iteration solution. Motivated by the property, the minimization problem (39) can be converted into the following relaxed problem by

$$\begin{aligned}
\mathbf{Z}_{G_k}^{t+1} &= \arg\min_{\mathbf{X}_{G_k}} \frac{1}{2} \left\| \mathbf{R}_{G_k} - \mathbf{Z}_{G_k} \right\|_2^2 + \tau \Re(\mathbf{Z}_{G_k}) \\
&= \arg\min_{\mathbf{X}_{G_k}} \frac{1}{2} \left\| \mathbf{R}_{G_k} - \mathbf{Z}_{G_k} \right\|_2^2 + \tau \sum_{i=1}^{r} \rho\left(\sigma_i(\mathbf{Z}_{G_k})\right) \\
&\approx \arg\min_{\mathbf{X}_{G_k}} \frac{1}{2} \left\| \mathbf{R}_{G_k} - \mathbf{Z}_{G_k} \right\|_2^2 + \tau \sum_{i=1}^{r} \left[ \rho\left(\sigma_i(\mathbf{Z}_{G_k}^t)\right) + \widetilde{\omega}_{k,i}^t \left(\sigma_i(\mathbf{Z}_{G_k}) - \sigma_i(\mathbf{Z}_{G_k}^t)\right) \right] \\
&= \arg\min_{\mathbf{X}_{G_k}} \frac{1}{2} \left\| \mathbf{R}_{G_k} - \mathbf{Z}_{G_k} \right\|_2^2 + \tau \sum_{i=1}^{r} \left( \widetilde{\omega}_{k,i}^t \sigma_i(\mathbf{Z}_{G_k}) \right)
\end{aligned} \tag{43}$$

where $\widetilde{\omega}_{k,i}^t \in \partial \left[ \rho\left(\sigma_i(\mathbf{Z}_{G_k}^t)\right) \right]$ denotes the $i$-th reweighting for each singular value.

**Theorem 3.2.** [55][56] *For any $\lambda > 0$, $\mathbf{R} \in \mathbb{R}^{M \times N}$, and the weighting vector $\widetilde{\boldsymbol{\omega}} = [\widetilde{\omega}_1, \widetilde{\omega}_2, \cdots, \widetilde{\omega}_s]$ with $0 \leq \widetilde{\omega}_1 \leq \widetilde{\omega}_2 \leq \cdots \leq \widetilde{\omega}_s, (s = \min(M, N))$, then the globally optimal solution of following optimization problem*

$$\min \frac{1}{2} \|\mathbf{R} - \mathbf{Z}\|_F^2 + \lambda \sum_{i=1}^{s} w_i \sigma_i(\mathbf{Z}) \tag{44}$$

*is given by the weighted singular value thresholding (WSVT)*

$$\mathbf{Z}^* = \mathbf{U} S_{\lambda \widetilde{\boldsymbol{\omega}}}(\boldsymbol{\Sigma}) \mathbf{V}^T \tag{45}$$

*where $\mathbf{R} = \mathbf{U} \boldsymbol{\Sigma} \mathbf{V}^T$ denotes the SVD of $\mathbf{R}$, and for each diagonal element $\boldsymbol{\Sigma}_{ii}$ of $\boldsymbol{\Sigma}$, there is*

$$S_{\lambda \widetilde{\boldsymbol{\omega}}}(\boldsymbol{\Sigma}) = \text{Diag}\{(\boldsymbol{\Sigma}_{ii} - \lambda w_i)_+\}. \tag{46}$$

According to the nonnegativity and the monotonicity of $\partial \rho(\theta)$ and $\omega_{k,i}$, we can easily obtain the following relationship between $\widetilde{\omega}_{k,i}^t, i = 1,2,\cdots r$, that is $0 \leq \widetilde{\omega}_{k,1}^t \leq \widetilde{\omega}_{k,2}^t \leq \cdots \leq \widetilde{\omega}_{k,r}^t$. Hence, according to the **Theorem 3.2**, the globally optimal solution of (43) can be given by the following WSVT operator

$$\mathbf{Z}_{G_k}^{t+1} = \mathbf{U}_{G_k} S_{\tau w_{k,i}^t}(\boldsymbol{\Sigma}_{G_k}) \mathbf{V}_{G_k}^T \tag{47}$$

where $S_{\tau w_{k,i}^t}(\boldsymbol{\Sigma}_{G_k}) = \text{Diag}\left\{ \left((\boldsymbol{\Sigma}_{G_k})_{ii} - \tau \widetilde{\omega}_{k,i}^t\right)_+ \right\}$, in which $(z)_+ = \max\{z, 0\}$, $\mathbf{U}_{G_k}$, $\boldsymbol{\Sigma}_{G_k} = diag(\varsigma_{k,1}, \varsigma_{k,2}, \cdots, \varsigma_{k,r})$ and $\mathbf{V}_{G_k}^T$ are achieved by the singular value decomposition (SVD) of $\mathbf{R}_{G_k}$, e.g., $\mathbf{R}_{G_k} = \mathbf{U}_{G_k} \boldsymbol{\Sigma}_{G_k} \mathbf{V}_{G_k}^T$, $(\boldsymbol{\Sigma}_{G_k})_{ii}$ denotes the $i$-th singular value $\varsigma_{k,i}$. By iteratively updating the reweighting $\widetilde{\omega}_{k,i}^t \in \partial \left[ g\left(\sigma_i(\mathbf{Z}_{G_k}^t)\right) \right]$, the problem (43) can be resolved effectively. The whole procedure of GSR-IRNN for denoising model (39) can be summarized in the **Algorithm 1**.

**Algorithm 1.** Proposed GSR-IRNN Algorithm for GSR based denoising model

$$\mathbf{Z}_{G_k}^{(t+1)} = \arg\min_{\mathbf{Z}_{G_k}} \frac{1}{2}\|\mathbf{R}_{G_k} - \mathbf{Z}_{G_k}\|_F^2 + \tau \sum_{i=1}^{r} \rho\left(\sigma_i(\mathbf{Z}_{G_k})\right)$$

**Input:** $\mathbf{R}_{G_k}$;

**Initialization:** $\tau$, $t = 0$, $\omega_i^{(0)}, i = 1,2,r$, $\mathbf{W}^{(0)} = \mathbf{0}$, $\mathbf{Z}^{(0)} = \mathbf{0}$;

**While** not convergence **do**

    Updating $\mathbf{Z}^{(t+1)}$ using the Eq. (47);

    Updating weightings by $\omega_i^{(t+1)} \in \partial\left[g\left(\sigma_i(\mathbf{Z}_{G_k}^{t+1})\right)\right]$;

    $t = t + 1$;

**End while.**

**Output:** The reconstructed image group $\mathbf{Z}_{G_k}^{(t+1)}$.

### 3.4 Iteratively-weighting strategy

Although the proposed nonconvex model (16) can improve the NNM effectively, it still has some problems. According to the theory of low rank minimization, the rank of a certain matrix only corresponds to the larger nonzero singular values, what's more, larger singular values often contain more information of matrix. For better approximation of the rank of group matrix, hence, the larger singular values should be shrunk less, and the smaller ones should be shrunk more. In this paper, we propose a more flexible and effective iteratively-weighted strategy for corresponding singular value $\sigma_i(\mathbf{Z}_{G_k})$ to avoid over-shrinking. Our motivation is to shrink the larger singular values less and shrink the smaller ones more. intuitively, each weighting should be inversely proportional to $|\sigma_i(\mathbf{Z}_{G_k}^t)|$, it can be expressed as

$$\overline{\omega}_{k,i}^t = \frac{1}{|\sigma_i(\mathbf{Z}_{G_k}^t)| + \varepsilon} \tag{48}$$

where $\sigma_i(\mathbf{Z}_{G_k}^t), i = 1,2,\cdots,min(B_s,c)$ denotes the $i$-th singular value of $t$-th iteration solution $\mathbf{Z}_{G_k}^t$, and the small constant parameter $\varepsilon$ can prevent the denominator from zero, e.g., $2.2204e^{-16}$. Accordingly, the optimization of (43) can be converted the following weighting model, e.g.,

$$\mathbf{Z}_{G_k}^{t+1} = \arg\min_{\mathbf{X}_{G_k}} \frac{1}{2}\|\mathbf{R}_{G_k} - \mathbf{Z}_{G_k}\|_2^2 + \lambda \sum_{i=1}^{r}\left(w_{k,i}^t \rho\left(\sigma_i(\mathbf{Z}_{G_k})\right)\right) \tag{49}$$

where $w_{k,i}^t = \overline{\omega}_{k,i}^t \widetilde{\omega}_{k,i}^t = \partial\rho\left(\sigma_i(\mathbf{Z}_{G_k}^t)\right)/(|\sigma_i(\mathbf{Z}_{G_k}^t)| + \varepsilon)$.

According to the nonnegativity and the monotonicity of $\partial\rho(\theta)$ and $\overline{\omega}_{k,i}$, we can easily obtain the following relationship between $w_{k,i}, i = 1,2,\cdots r$, that is $0 \le w_{k,1} \le w_{k,2} \le \cdots \le w_{k,r}$. Hence, according to the **Theorem 3.2**, the globally optimal solution of (49) can be given by

$$\mathbf{Z}_{G_k}^{t+1} = \mathbf{U}_{G_k} S_{\tau w_{k,i}^t}(\mathbf{\Sigma}_{G_k})\mathbf{V}_{G_k}^{\mathrm{T}} \tag{50}$$

where $S_{\tau w_{k,i}^t}(\Sigma_{G_k}) = \text{Diag}\left\{\left((\Sigma_{G_k})_{ii} - \tau w_{k,i}^t\right)_+\right\}$, in which $(x)_+ = \max\{x, 0\}$, $\mathbf{U}_{G_k}$, $\Sigma_{G_k} = diag(\varsigma_{k,1}, \varsigma_{k,2}, \cdots, \varsigma_{k,r})$ and $\mathbf{V}_{G_k}^T$ are achieved by the singular value decomposition (SVD) of $\mathbf{R}_{G_k}$, e.g., $\mathbf{R}_{G_k} = \mathbf{U}_{G_k}\Sigma_{G_k}\mathbf{V}_{G_k}^T$, $(\Sigma_{G_k})_{ii}$ denotes the $i$-th singular value $\varsigma_{k,i}$. By iteratively updating the reweighting $\widetilde{\omega}_{k,i}^t \in \partial\left[g\left(\sigma_i(\mathbf{Z}_{G_k}^t)\right)\right]$ and $\overline{\omega}_{k,i}^t = 1/(|\sigma_i(\mathbf{Z}_{G_k}^t)| + \varepsilon)$, the problem (45) can be resolved effectively.

After achieving all $\mathbf{Z}_{G_k}^t, k = 1,2,\cdots,n$, then we can obtain the solution of $\mathbf{Z}$-subproblem by averaging all $\mathbf{Z}_{G_k}^t$

$$\mathbf{Z}^{t+1} = \sum_{k=1}^n G_k^T(\mathbf{Z}_{G_k}^t)./\sum_{k=1}^n G_k^T(\mathbf{1}_{\mathcal{B}_c}) \tag{51}$$

where the $G_k^T(\cdot)$ denotes the transpose grouping operator, which can reconstruct the original image from the group.

### 3.4 Summary of the GSR-AIR Algorithm

The whole procedure of our proposed algorithm of GSR-AIR can be shown in the **Algorithm 1 and Algorithm 2** for standard CS and robust CS respectively.

---

**Algorithm 2.** Proposed GSR-AIR Algorithm for Standard image CS

$$\widehat{\mathbf{X}} = \arg\min_{\mathbf{X}} \frac{1}{2}\|\mathbf{Y} - \mathbf{H}\mathbf{X}\|_2^2 + \lambda \sum_{k=1}^n \Re(\mathbf{X}_{G_k})$$

---

**Input:** The Observation $\mathbf{Y}$, the compressed sampling matrix $\mathbf{H}$;

**Initialization:** $c$, $\mathcal{B}_s$, $t = 0$, $\delta$, $\varepsilon$, $\mu$, $\gamma$, $\lambda^{(0)}$, $\mathbf{W}^{(0)} = \mathbf{0}$, $\mathbf{Z}^{(0)} = \mathbf{0}$;

**Repeat**

  Updating $\mathbf{X}^{(t+1)}$ using the Eq. (22) (23);
  Computing $\mathbf{R}^{(t+1)} = \mathbf{X}^{(t+1)} - \mathbf{W}^{(t)}$;

  Constructing groups $\{\mathbf{R}_{G_k}^{(t+1)}\}$ from $\mathbf{R}^{(t+1)}$;

  Computing parameter $\tau = \frac{\lambda K}{\mu N}$;

  for each group $\mathbf{R}_{G_k}^{(t+1)}$

    Singular value decomposition by $\mathbf{R}_{G_k}^{(t+1)} = \mathbf{U}_{G_k}\Sigma_{G_k}\mathbf{V}_{G_k}^T$;

    Updating weightings $\overline{\omega}_i^{(t+1)}$ and $\widetilde{\omega}_{k,i}^{(t+1)}$;

    Reconstruct $\mathbf{Z}_{G_k}^{(t+1)}$ using the Eq. (50);

    Computing $\mathbf{Z}^{(t+1)}$ by averaging all $\mathbf{Z}_{G_k}^{(t+1)}$ using the Eq. (51);

  end
  Computing $\mathbf{W}^{(t+1)}$ using Eq. (20);
  $t = t + 1$;

**Until the maximum iteration number is reached.**

---

**Output:** The reconstructed image $\mathbf{X}^{(t+1)}$.

---

**Algorithm 3.** Proposed GSR-AIR Algorithm for Robust image CS

$$\widehat{\mathbf{X}} = \arg\min_{\mathbf{X}} \frac{1}{2}\psi(\mathbf{Y} - \mathbf{H}\mathbf{X}) + \lambda \sum_{k=1}^{n} \Re(\mathbf{X}_{G_k})$$

**Input:** The Observation $\mathbf{Y}$, the compressed sampling matrix $\mathbf{H}$;

**Initialization:** $c$, $\mathcal{B}_s$, $t = 0$, $\delta$, $\varepsilon$, $\mu$, $\gamma$, $\lambda^{(0)}$, $\mathbf{W}^{(0)} = \mathbf{0}$, $\mathbf{Z}^{(0)} = \mathbf{0}$;

**Repeat**

    Updating $\mathbf{X}^{(t+1)}$ using the Eq. (34) (35);
    Computing $\mathbf{R}^{(t+1)} = \mathbf{X}^{(t+1)} - \mathbf{W}^{(t)}$;

    Constructing groups $\{\mathbf{R}_{G_k}^{(t+1)}\}$ from $\mathbf{R}^{(t+1)}$;

    Computing parameter $\tau = \frac{\lambda K}{\mu N}$;

    for each group $\mathbf{R}_{G_k}^{(t+1)}$

        *Singular value decomposition by* $\mathbf{R}_{G_k}^{(t+1)} = \mathbf{U}_{G_k} \mathbf{\Sigma}_{G_k} \mathbf{V}_{G_k}^T$;

        *Updating weightings* $\bar{\omega}_i^{(t+1)}$ *and* $\widetilde{\omega}_{k,i}^{(t+1)}$;

        *Reconstruct* $\mathbf{Z}_{G_k}^{(t+1)}$ *using the Eq.* (50);

        *Computing* $\mathbf{Z}^{(t+1)}$ *by averaging all* $\mathbf{Z}_{G_k}^{(t+1)}$ *using the Eq.* (51);

    end
    Computing $\mathbf{W}^{(t+1)}$ using Eq. (20);
    $t = t + 1$;

**Until the maximum iteration number is reached.**

**Output:** The reconstructed image $\mathbf{X}^{(t+1)}$.

## IV. Experimental Results

In this section, we employ several classical nonconvex functions as surrogates to evaluate the performance of our proposed nonconvex framework for CS reconstruction problems, including the Lp, MCP, ETP and logarithm function. Because the original image $\mathbf{X}$ is unknown, this paper employs the result of a state-of-the art algorithm MH [57] as initialization for our proposed standard CS framework, and employs the result of DCT [37] as initialization for our robust CS framework. For MH initialization-based framework, we generate CS measurements by randomly sampling the image block, while for DCT initialization-based framework, the CS measurements are generated by randomly sampling the Fourier transform coefficients of the original images. To better illustrate the performance, we compare the performance of proposed algorithm with several state-of-the-art convex and nonconvex regularization CS reconstruction methods. We also analyze the convergence behavior of our proposed GSR-AIR algorithm for our proposed nonconvex optimization model. We introduce two metrics to evaluate the reconstruction performance of all algorithms, namely, the peak signal-to-noise ratio (PSNR) and the metric feature similarity (FSIM) [58]. All the natural images for experiments are listed in the **Fig**. 3.

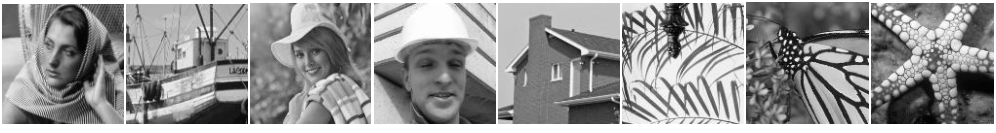

**Fig. 3**. Eight typical $256 \times 256$ natural images for experiments: Barbara, Boats, Elaine, Foreman, House, Leaves, Monarch and Starfish.

## 4.1 Standard Image CS
### 4.1.1 Parameters setting

From the **Algorithm 2**, we can find that there are some important parameters, the regularization parameter $\lambda$, the penalty factor $\mu$, and the optimal number of similar patch $c$, and other parameters, such as the patch size of $\sqrt{B_s} \times \sqrt{B_s}$. Empirically, in this paper, we will set the block size $\sqrt{B_s} \times \sqrt{B_s}$ as $32 \times 32$, and the patch size is set to be $6 \times 6$, and the searching window $L \times L$ is set to be $20 \times 20$ for all the experiments. To evaluate the effects of other two parameters for the reconstructed quality and choose optimal parameters, in this subsection, we plot the PSNRs curve and FSIMs curve for our proposed algorithm versus $\mu$ and $c$ respectively. For the penalty factor parameter, we empirically choose $\mu \in [10^{-6}, 1]$, where we fix other parameters and then evaluate the PSNRs and FSIMs. We choose the typical image 'House' to carry out the experiments under three different sampling rates of 0.2, 0.3 and 0.4. **Fig**. 4 (a) and (b) present the PSNRs curve and FSIMs curve respectively, from the results we can observe that our proposed algorithm can obtain the best reconstruction quality when $\mu \in (10^{-3} \sim 10^{-1})$ for all the measurements.

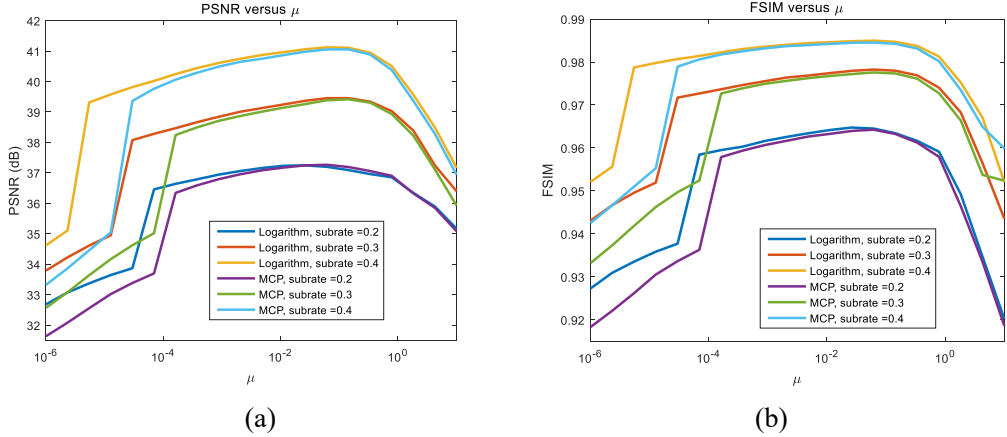

(a)                                        (b)

**Fig. 4**. Performance of CS reconstruction for 0.1, 0.2 and 0.3 measurements of 'House'. (a) PSNR versus the parameter $\mu$; (b) FSIM versus the parameter $\mu$.

For the best match parameter $c$, we conduct experiments on several typical images by employing two typical surrogate functions of Logarithm and SCAD. We fix sampling rates to be 0.2, 0.3 and 0.4. **Fig**. 5 (a) and (b) plot the PSNRs curve and the FSIMs curve versus $c$, from the results we can also observe that the proposed algorithm is not sensitive to the parameter $c$, and can achieve the favorable PSNRs and FSIMs when the $c \in [60, 100]$, however, we empirically find that a larger $c$ will bring higher computational time cost. Hence, in this paper, we set $c = 60$ for all of our following experiments. The selections of all parameters are detailed in the table I for different sub-sampling rates and surrogate functions.

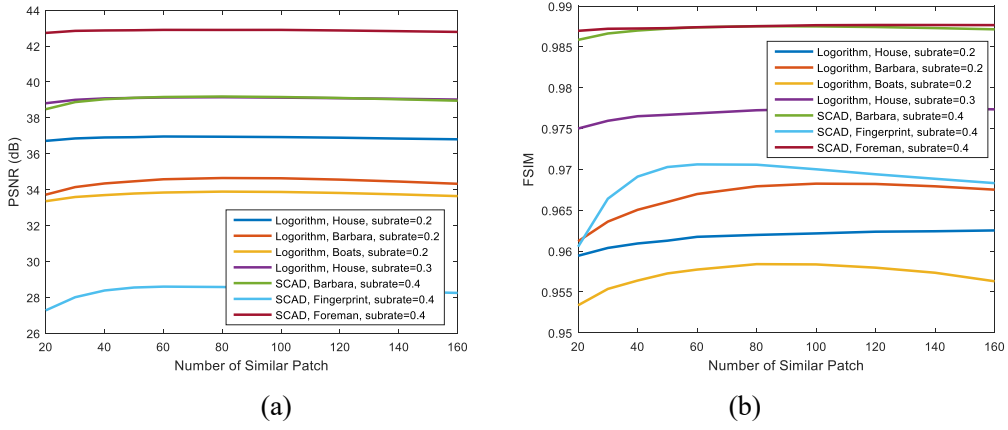

**Fig. 5**. Performance of CS reconstruction for three different images of 'House', 'Barbara' and 'Boats'. (a) PSNR versus the number of similar patch $c$; (b) FSIM versus the number of similar patch $c$.

**4.1.2 Effect comparison with the nonconvex weighting strategy**

(I) Effect comparison with the nonconvex strategy

To evaluate the effect of our proposed nonconvex strategy, this subsection will conduct comparative experiments between the convex weighted model (Weighted NNM) and our proposed nonconvex weighted model, we employ the popular functions of ETP, Logarithm, Lp and MCP as the case and reconstructing four images from under-sampled data with different sub-rates. **Fig**. 6 presents the PSNR curves comparisons of our proposed method using Logarithm and weighted NNM versus the iteration number for the image 'Barbara' case, and all the comparable results are listed in the Table II, we can find that our proposed nonconvex strategy can improve the corresponding weighted NNM based model significantly, particularly for lower under-sampling rate, e,g., 0.1. It should be noted that all parameters for the competing convex NNM algorithm are setting so that it can achieve best performance for fair comparisons, e.g., $\mu = 0.0001, \lambda = 0.01$ for 10% under-sampled data, $\mu = 0.0005, \lambda = 0.01$ for 20% under-sampled data, and $\mu = 0.0001, \lambda = 0.01$ for 30% under-sampled data.

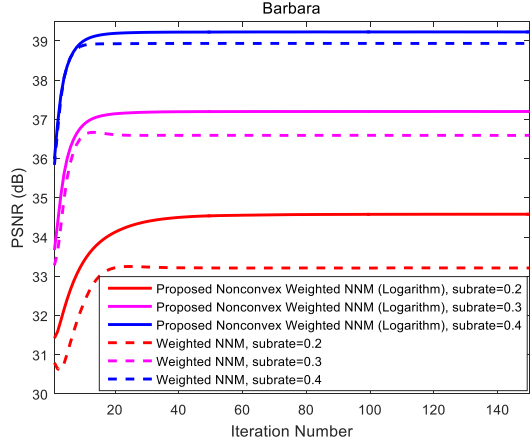

**Fig. 6**. PSNR curves comparisons versus the iteration number between the weighted NNM and our proposed nonconvex weighted NNM model (Logarithm).

**Table II**  PSNR (dB)/FSIM comparisons of proposed nonconvex method with the traditional weighted convex NNM

| Rate | Method | *Barbara* | *Boats* | *Foreman* | *House* |
|---|---|---|---|---|---|
| 10% | Weighted NNM | 28.41/0.9079 | 28.24/0.8977 | 35.45/0.9465 | 33.29/0.9270 |
|  | **Proposed (ETP)** | **29.82/0.9244** | **29.03/0.9076** | **35.96/0.9492** | **34.16/0.9300** |
|  | **Proposed (logarithm)** | **29.91/0.9264** | **29.22/0.9098** | **36.16/0.9510** | **34.24/0.9310** |
|  | **Proposed (Lp)** | **29.95/0.9270** | **29.24/0.9102** | **36.20/0.9513** | **34.26/0.9315** |
|  | **Proposed (MCP)** | **29.90/0.9264** | **29.31/0.9105** | **36.21/0.9510** | **34.29/0.9316** |
| 20% | Weighted NNM | 34.16/0.9647 | 33.59/0.9562 | 38.63/0.9701 | 36.92/0.9616 |
|  | **Proposed (ETP)** | **34.67/0.9675** | **33.96/0.9586** | **38.75/0.9703** | **37.08/0.9623** |
|  | **Proposed (logarithm)** | **34.36/0.9656** | **33.73/0.9567** | **38.69/0.9701** | **36.97/0.9615** |
|  | **Proposed (Lp)** | **34.38/0.9658** | **33.77/0.9572** | **38.71/0.9702** | **37.00/0.9618** |
|  | **Proposed (MCP)** | **34.46/0.9665** | **33.91/0.9584** | **38.78/0.9702** | **37.13/0.9626** |
| 30% | Weighted NNM | 36.90/0.9799 | 36.72/0.9760 | 40.87/0.9810 | 38.99/0.9763 |
|  | **Proposed (ETP)** | **37.29/0.9814** | **37.22/0.9777** | **41.13/0.9816** | **39.36/0.9780** |
|  | **Proposed (logarithm)** | **37.16/0.9808** | **37.03/0.9770** | **41.02/0.9812** | **39.15/0.9768** |
|  | **Proposed (Lp)** | **37.18/0.9809** | **37.09/0.9773** | **41.05/0.9813** | **39.23/0.9773** |
|  | **Proposed (MCP)** | 36.87/**0.9797** | 36.68/**0.9757** | 40.89/**0.9808** | 38.97/**0.9758** |

(II) Effect comparison with the weighting strategy

To evaluate the effect of our proposed weighting strategy, this subsection will conduct comparative experiments between the corresponding nonconvex NNM model without weighting strategy and our proposed nonconvex weighted model, we reconstruct four different images from under-sampled data using four surrogate functions of ETP, Logarithm, Lp and MCP. It should be noted that the non-weighted Lp based GSR method is a recent proposed method for CS problem [45]. **Fig**. 7 presents the PSNR curves comparisons versus the iteration number for the image 'House' case, and Table IV presents all comparable results. From them we can find that our proposed weighting strategy can outperform the corresponding nonconvex NNM based model without weighting strategy significantly. It should be noted that all parameters for the competing convex

NNM algorithm are setting for the best performance for a for a fair comparison, which are listed in the table III.

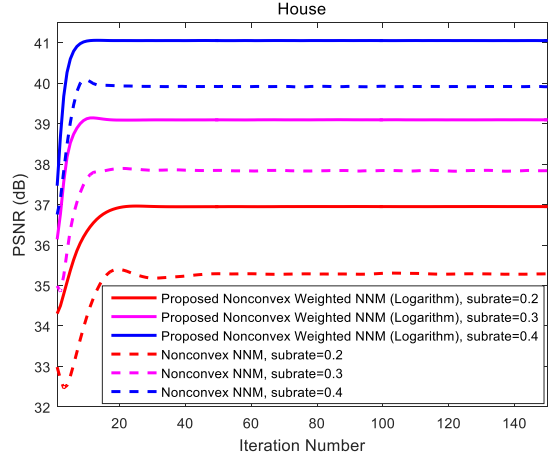

**Fig. 7**. PSNR curves comparisons versus the iteration number between the nonconvex NNM and our proposed nonconvex weighted NNM model (Logarithm).

**Table IV**   PSNR/FSIM comparisons of proposed nonconvex weighted method with the nonconvex method without weighting

| Rate | Method | *Elaine* | *Leaves* | *Monarch* | *Starfish* |
|---|---|---|---|---|---|
| 10% | Nonconvex NNM (ETP) | 32.04/0.9277 | **25.86/0.9139** | **27.39/0.9068** | 25.62/0.9277 |
|  | **Proposed 1 (ETP)** | **32.25/0.9300** | 25.71/**0.9141** | 27.29/0.9066 | **25.43/0.8775** |
|  | Nonconvex NNM (Logarithm) | 31.45/0.9234 | 25.54/0.9098 | 26.93/0.9028 | 25.22/0.8694 |
|  | **Proposed 1 (Logarithm)** | **32.36/0.9319** | **25.84/0.9150** | **27.42/0.9088** | **25.51/0.8792** |
|  | Nonconvex NNM (Lp) | 31.29/0.9230 | 25.17/0.9063 | 26.62/0.9004 | 24.97/0.8651 |
|  | **Proposed 1 (Lp)** | **32.38/0.9324** | **25.87/0.9151** | **27.45/0.9092** | **25.52/0.8795** |
|  | Nonconvex NNM (MCP) | 30.42/0.9150 | 23.88/0.8887 | 25.54/0.8861 | 24.14/0.8466 |
|  | **Proposed 1 (MCP)** | **32.36/0.9324** | **25.89/0.9144** | **27.47/0.9095** | **25.52/0.8788** |
| 20% | Nonconvex NNM (ETP) | 36.19/0.9650 | 31.48/**0.9620** | 31.81/0.9505 | 29.92/0.9404 |
|  | **Proposed 1 (ETP)** | **36.15**/0.9644 | **31.48**/0.9618 | **31.84/0.9508** | **30.01/0.9406** |
|  | Nonconvex NNM (Logarithm) | 35.76/0.9627 | 30.95/0.9691 | 31.39/0.9481 | 29.38/0.9342 |
|  | **Proposed 1 (Logarithm)** | **35.95/0.9632** | **31.45/0.9611** | **31.74/0.9503** | **29.87/0.9385** |
|  | Nonconvex NNM (Lp) | 35.72/0.9628 | 30.57/0.9577 | 31.18/0.9467 | 29.20/0.9329 |
|  | **Proposed 1 (Lp)** | **35.98/0.9635** | **31.43/0.9612** | **31.74/0.9503** | **29.86/0.9386** |
|  | Nonconvex NNM (MCP) | 34.54/0.9554 | 27.78/0.9375 | 29.30/0.9317 | 27.40/0.9095 |
|  | **Proposed 1 (MCP)** | **36.08/0.9644** | **31.42/0.9616** | **31.76/0.9502** | **29.83/0.9392** |
| 30% | Nonconvex NNM (ETP) | 38.19/0.9765 | 35.11/0.9795 | 34.80/0.9677 | 33.17/0.9640 |
|  | **Proposed 1 (ETP)** | **38.33/0.9773** | **35.24/0.9803** | **34.85**/0.9671 | **33.43/0.9660** |
|  | Nonconvex NNM (Logarithm) | 37.85/0.9751 | 34.25/0.9767 | 34.18/0.9654 | 32.39/0.9594 |
|  | **Proposed 1 (Logarithm)** | **38.25/0.9768** | **35.20/0.9799** | **34.89/0.9678** | **33.31/0.9649** |
|  | Nonconvex NNM (Lp) | 37.42/0.9733 | 33.17/0.9725 | 33.39/0.9619 | 31.59/0.9539 |
|  | **Proposed 1 (Lp)** | **38.27/0.9769** | **35.23/0.9801** | **34.90/0.9678** | **33.33/0.9653** |

| | | | | |
|---|---|---|---|---|
| Nonconvex NNM (MCP) | 36.72/0.9705 | 31.02/0.9619 | 31.96/0.9533 | 30.09/0.9424 |
| **Proposed 1 (MCP)** | **38.09/0.9760** | **34.87/0.9788** | **34.63/0.9671** | **32.95/0.9627** |

### 4.1.3 Convergence analysis

Although the nonconvex penalized regularization model can obtain better performance than the convex surrogate, it is intractable to demonstrate the convergence of our proposed algorithm. In this subsection, we will present the convergence property of our proposed algorithm visually by the PSNR curve versus the iteration number. **Fig**. 8 (a), (b) and (c) present the PSNRs curves for Logarithm function, MCP function and SCAD function under different sub-sampling rates, from the results we can observe that our proposed algorithm for the nonconvex framework contains good convergence property.

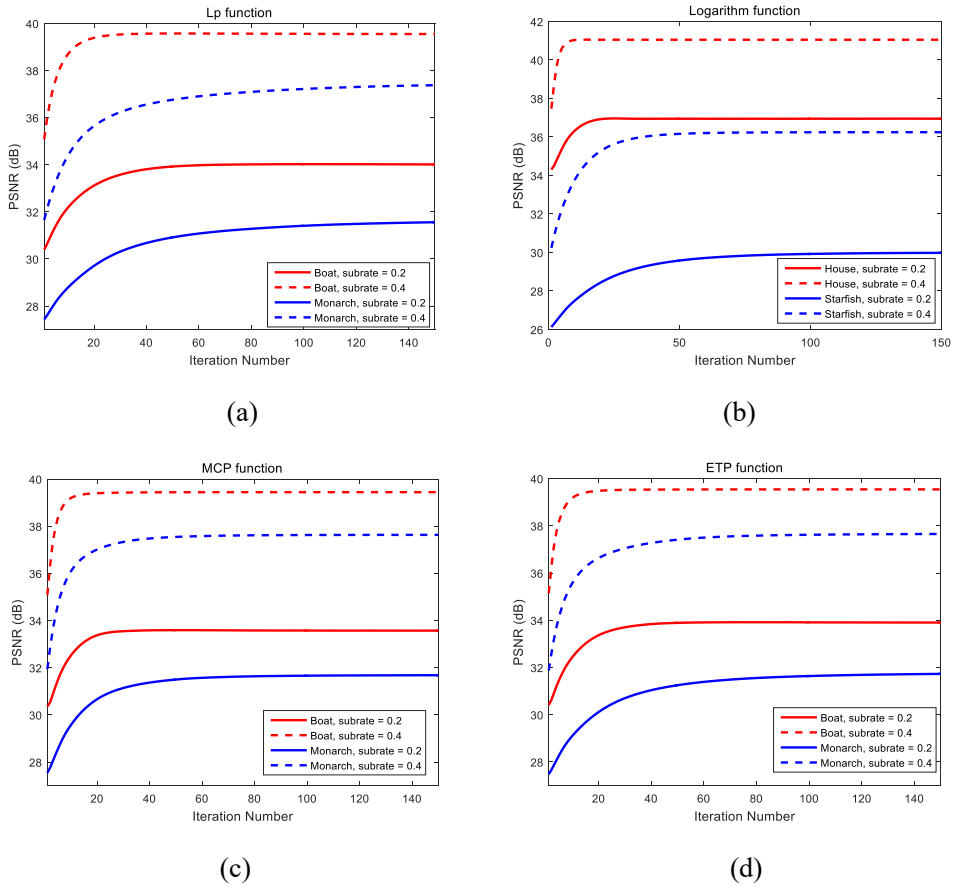

**Fig. 8**. The convergence of the proposed algorithm for Logarithm function, MCP function and SCAD function with different rates of 0.2 and 0.4.

### 4.1.4 Comparisons with state-of-the art approaches

To demonstrate the effectiveness of our proposed algorithm, we employ six representative convex CS recovery algorithms for comparisons, including the algorithms of MH [57], SGSR [59], ALSB [60], JASR [61], GSR-Lp [45], GSR-NCR [46]. It should be noted that all these competing methods employ the results of MH as initializations. Table V and VI list the results of PSNR and FSIM of six state-of-the art algorithms and our proposed nonconvex algorithm under five different sampling rates of 0.1, 0.2, 0.3, 0.4 and 0.5. We can observe that our nonconvex model can obtain higher PSNR values and FSIM values in lower sampling rates, e.g., 0.1 and 0.2.

**Table V** The PSNR comparisons of proposed algorithm and other state-of-the art algorithms

| Rate | Method | *Barbara* | *Boats* | *Elaine* | *Foreman* | *House* | *Leaves* | *Monarch* | *Starfish* | *Average* |
|---|---|---|---|---|---|---|---|---|---|---|
| 10% | MH based recovery | 26.74 | 26.09 | 29.36 | 33.14 | 30.32 | 20.90 | 23.20 | 22.53 | 26.54 |
| | SGSR | 28.70 | 27.71 | 31.32 | 34.88 | 32.77 | 22.22 | 24.27 | 22.91 | 28.10 |
| | ALSB | 27.01 | 27.75 | 30.99 | 33.49 | 32.18 | 21.37 | 24.27 | 23.63 | 27.59 |
| | JASR | 29.58 | 28.59 | 32.01 | 35.61 | 33.49 | 23.62 | 25.83 | 24.39 | 29.14 |
| | GSR-Lp | 28.38 | 28.37 | 31.27 | 35.57 | 33.46 | 25.17 | 26.61 | 24.96 | 29.22 |
| | GSR-NCR | 28.28 | 27.62 | 31.35 | 35.59 | 32.35 | 21.74 | 23.86 | 22.92 | 27.96 |
| | Proposed 1 (ETP) | **29.82** | **29.03** | **32.25** | **35.96** | **34.16** | **25.71** | **27.29** | **25.43** | **29.96** |
| | Proposed 1 (logarithm) | **29.91** | **29.22** | **32.36** | **36.16** | **34.24** | **25.84** | **27.42** | **25.51** | **30.08** |
| | Proposed 1 (Lp) | **29.95** | **29.24** | **32.38** | **36.20** | **34.26** | **25.87** | **27.45** | **25.52** | **30.11** |
| | Proposed 1 (MCP) | **29.90** | **29.31** | **32.36** | **36.21** | **34.29** | **25.89** | **27.47** | **25.52** | **30.12** |
| 20% | MH based recovery | 30.81 | 29.92 | 33.47 | 35.92 | 33.85 | 25.16 | 27.11 | 25.92 | 30.27 |
| | SGSR | 33.45 | 32.41 | 34.86 | 36.98 | 35.81 | 28.74 | 28.76 | 27.19 | 32.28 |
| | ALSB | 31.77 | 33.04 | 35.11 | 35.33 | 35.93 | 27.14 | 28.39 | 27.20 | 31.74 |
| | JASR | 34.16 | 33.21 | 35.66 | 37.87 | 36.10 | 30.24 | 30.60 | 29.10 | 33.37 |
| | GSR-Lp | 33.74 | 33.34 | 35.72 | 38.65 | 37.02 | 30.33 | 31.04 | 29.01 | 33.61 |
| | GSR-NCR | 33.91 | 33.30 | 35.61 | 37.74 | 36.57 | 28.89 | 29.41 | 27.88 | 32.91 |
| | Proposed 1 (ETP) | **34.67** | **33.96** | **36.15** | **38.75** | **37.08** | **31.48** | **31.84** | **30.01** | **34.24** |
| | Proposed 1 (logarithm) | **34.36** | **33.73** | **35.95** | **38.69** | **36.97** | **31.45** | **31.74** | **29.87** | **34.10** |
| | Proposed 1 (Lp) | **34.38** | **33.77** | **35.98** | **38.71** | **37.00** | **31.43** | **31.74** | **29.86** | **34.11** |
| | Proposed 1 (MCP) | **34.46** | **33.91** | **36.08** | **38.78** | **37.13** | **31.42** | **31.76** | **29.83** | **34.17** |
| 30% | MH based recovery | 32.99 | 32.26 | 35.40 | 37.69 | 35.69 | 27.65 | 29.21 | 27.88 | 32.35 |
| | SGSR | 35.91 | 35.22 | 36.87 | 38.47 | 37.37 | 32.98 | 31.99 | 30.79 | 34.95 |
| | ALSB | 34.70 | 36.45 | 37.49 | 36.50 | 38.36 | 31.30 | 31.37 | 30.43 | 34.58 |
| | JASR | 36.59 | 36.08 | 36.83 | 38.54 | 38.04 | 33.70 | 33.63 | 32.33 | 35.72 |
| | GSR-Lp | 35.67 | 35.30 | 37.39 | 40.34 | 38.32 | 33.17 | 33.39 | 31.59 | 35.65 |
| | GSR-NCR | 37.16 | **37.26** | 38.25 | **41.18** | 39.37 | 34.92 | 34.64 | 33.17 | 36.99 |
| | Proposed 1 (ETP) | **37.29** | 37.22 | **38.33** | 41.13 | 39.36 | **35.24** | 34.85 | 33.43 | **37.11** |
| | Proposed 1 (logarithm) | 37.16 | 37.03 | **38.25** | 41.02 | 39.15 | **35.20** | 34.89 | 33.31 | **37.01** |
| | Proposed 1 (Lp) | **37.18** | 37.09 | **38.27** | 41.05 | 39.23 | **35.23** | 34.90 | 33.33 | **37.04** |
| | Proposed 1 (MCP) | 36.87 | 36.68 | 38.09 | 40.89 | 38.97 | 34.87 | 34.63 | 32.95 | 36.74 |
| 40% | MH based recovery | 35.13 | 34.22 | 37.07 | 39.15 | 36.64 | 29.68 | 31.14 | 29.60 | 34.08 |
| | SGSR | 37.70 | 37.41 | 38.63 | 39.84 | 38.99 | 35.83 | 34.66 | 33.66 | 37.09 |
| | ALSB | 37.23 | 38.88 | 39.48 | 42.62 | 40.06 | 34.47 | 34.52 | 33.24 | 37.56 |
| | JASR | 37.39 | 37.19 | 38.28 | 41.19 | 38.79 | 36.56 | 36.15 | 34.30 | 37.48 |
| | GSR-Lp | 38.33 | 38.43 | 39.59 | 42.42 | 40.56 | 36.86 | 36.47 | 34.73 | 38.42 |
| | GSR-NCR | **39.22** | **39.63** | 40.08 | 42.95 | **41.11** | 38.51 | 37.58 | 36.21 | 39.41 |
| | Proposed 1 (ETP) | 39.21 | 39.53 | **40.11** | 42.94 | 41.04 | **38.62** | 37.68 | **36.22** | **39.42** |
| | Proposed 1 (logarithm) | 39.13 | 39.43 | 40.06 | 42.86 | 40.97 | 38.49 | **37.60** | 36.08 | 39.33 |

|  | Method | Barbara | Boats | Elaine | Foreman | House | Leaves | Monarch | Starfish | Average |
|---|---|---|---|---|---|---|---|---|---|---|
|  | Proposed 1 (Lp) | 39.16 | 39.49 | **40.10** | 42.93 | 41.03 | **38.57** | **37.64** | 36.13 | 39.38 |
|  | Proposed 1 (MCP) | 39.15 | 39.49 | **40.10** | 42.97 | 41.08 | **38.54** | **37.58** | 36.08 | 39.37 |
| 50% | MH based recovery | 36.80 | 35.86 | 38.54 | 40.39 | 38.53 | 31.93 | 32.94 | 31.10 | 35.76 |
|  | SGSR | 39.38 | 39.30 | 40.08 | 41.15 | 40.56 | 38.19 | 36.99 | 35.88 | 38.94 |
|  | ALSB | 39.44 | 40.93 | 41.19 | 44.40 | 41.80 | 37.67 | 36.99 | 36.09 | 39.81 |
|  | JASR | 40.31 | 40.30 | 39.47 | 42.69 | 41.44 | 39.06 | 38.36 | 37.14 | 39.85 |
|  | GSR-Lp | 40.28 | 40.42 | 41.20 | 44.07 | 42.03 | 39.55 | 38.72 | 37.03 | 40.41 |
|  | GSR-NCR | 41.04 | 41.51 | 41.67 | 44.54 | 42.58 | 41.33 | 40.19 | 38.49 | 41.42 |
|  | Proposed 1 (ETP) | **41.08** | **41.54** | **41.72** | **44.93** | **42.74** | **41.73** | **40.24** | **38.57** | **41.57** |
|  | Proposed 1 (logarithm) | **41.09** | **41.56** | **41.73** | **44.91** | **42.73** | **41.72** | **40.25** | **38.58** | **41.57** |
|  | Proposed 1 (Lp) | **41.08** | **41.55** | **41.71** | **44.92** | **42.73** | **41.73** | **40.24** | **38.57** | **41.57** |
|  | Proposed 1 (MCP) | **41.08** | **41.55** | **41.75** | **44.84** | **42.69** | **41.58** | **40.17** | **38.50** | **41.52** |

**Table VI** The FSIM comparisons of proposed algorithm and other state-of-the art algorithms

| Rate | Method | Barbara | Boats | Elaine | Foreman | House | Leaves | Monarch | Starfish | Average |
|---|---|---|---|---|---|---|---|---|---|---|
| 10% | MH based recovery | 0.8911 | 0.8481 | 0.9018 | 0.9267 | 0.8936 | 0.7634 | 0.7912 | 0.8078 | 0.8530 |
|  | SGSR | 0.9147 | 0.8915 | 0.9220 | 0.9393 | 0.9187 | 0.8356 | 0.8371 | 0.8177 | 0.8846 |
|  | ALSB | 0.8903 | 0.8830 | 0.9184 | 0.9254 | 0.9069 | 0.7934 | 0.8218 | 0.8343 | 0.8717 |
|  | JASR | 0.9223 | 0.9035 | 0.9282 | 0.9437 | 0.9167 | 0.8799 | 0.8822 | 0.8516 | 0.9035 |
|  | GSR-Lp | 0.9062 | 0.8983 | 0.9229 | 0.9473 | 0.9269 | 0.9064 | 0.9003 | 0.8649 | 0.9092 |
|  | GSR-NCR | 0.9217 | 0.8977 | 0.9318 | 0.9449 | 0.9128 | 0.8367 | 0.8289 | 0.8227 | 0.8872 |
|  | Proposed 1 (ETP) | **0.9244** | **0.9076** | **0.9300** | **0.9492** | **0.9300** | **0.9141** | **0.9066** | **0.8775** | **0.9174** |
|  | Proposed 1 (logarithm) | **0.9264** | **0.9098** | **0.9319** | **0.9510** | **0.9310** | **0.9150** | **0.9088** | **0.8792** | **0.9191** |
|  | Proposed 1 (Lp) | **0.9270** | **0.9102** | **0.9324** | **0.9513** | **0.9315** | **0.9151** | **0.9092** | **0.8795** | **0.9195** |
|  | Proposed 1 (MCP) | **0.9264** | **0.9105** | **0.9324** | **0.9510** | **0.9316** | **0.9144** | **0.9095** | **0.8788** | **0.9193** |
| 20% | MH based recovery | 0.9393 | 0.9159 | 0.9452 | 0.9558 | 0.9389 | 0.8576 | 0.8751 | 0.8729 | 0.9126 |
|  | SGSR | 0.9615 | 0.9465 | 0.9551 | 0.9598 | 0.9502 | 0.9373 | 0.9132 | 0.8993 | 0.9403 |
|  | ALSB | 0.9501 | 0.9512 | 0.9597 | 0.9460 | 0.9541 | 0.9094 | 0.8965 | 0.8973 | 0.9330 |
|  | JASR | 0.9651 | 0.9521 | 0.9603 | 0.9636 | 0.9425 | 0.9516 | 0.9409 | 0.9295 | 0.9507 |
|  | GSR-Lp | 0.9627 | 0.9550 | 0.9628 | 0.9702 | 0.9630 | 0.9569 | 0.9453 | 0.9312 | 0.9559 |
|  | GSR-NCR | 0.9642 | 0.9526 | 0.9600 | 0.9578 | 0.9508 | 0.9415 | 0.9201 | 0.9158 | 0.9454 |
|  | Proposed 1 (ETP) | **0.9675** | **0.9586** | **0.9644** | **0.9703** | **0.9623** | **0.9618** | **0.9508** | **0.9406** | **0.9595** |
|  | Proposed 1 (logarithm) | **0.9656** | **0.9567** | **0.9632** | **0.9701** | **0.9615** | **0.9611** | **0.9503** | **0.9385** | **0.9584** |
|  | Proposed 1 (Lp) | **0.9658** | **0.9572** | **0.9635** | **0.9702** | **0.9618** | **0.9612** | **0.9503** | **0.9386** | **0.9586** |
|  | Proposed 1 (MCP) | **0.9665** | **0.9584** | **0.9644** | **0.9702** | **0.9626** | **0.9616** | **0.9502** | **0.9392** | **0.9591** |
| 30% | MH based recovery | 0.9588 | 0.9438 | 0.9614 | 0.9686 | 0.9569 | 0.8960 | 0.8990 | 0.9055 | 0.9362 |
|  | SGSR | 0.9762 | 0.9684 | 0.9695 | 0.9711 | 0.9648 | 0.9676 | 0.9469 | 0.9447 | 0.9637 |
|  | ALSB | 0.9716 | 0.9744 | 0.9742 | 0.9575 | 0.9730 | 0.9537 | 0.9296 | 0.9412 | 0.9594 |
|  | JASR | 0.9785 | 0.9723 | 0.9661 | 0.9649 | 0.9649 | 0.9719 | 0.9610 | 0.9580 | 0.9672 |
|  | GSR-Lp | 0.9749 | 0.9699 | 0.9731 | 0.9791 | 0.9729 | 0.9725 | 0.9619 | 0.9540 | 0.9698 |
|  | GSR-NCR | **0.9815** | **0.9783** | **0.9774** | 0.9829 | **0.9795** | 0.9799 | 0.9666 | 0.9654 | **0.9764** |
|  | Proposed 1 (ETP) | 0.9814 | 0.9777 | 0.9773 | **0.9816** | 0.9780 | **0.9803** | **0.9671** | **0.9660** | 0.9762 |

|  |  |  |  |  |  |  |  |  |  |
|---|---|---|---|---|---|---|---|---|---|
|  | Proposed 1 (logarithm) | 0.9808 | 0.9770 | 0.9768 | **0.9812** | 0.9768 | **0.9799** | **0.9678** | 0.9649 | 0.9757 |
|  | Proposed 1 (Lp) | 0.9809 | 0.9773 | 0.9769 | **0.9813** | 0.9773 | **0.9801** | **0.9678** | 0.9653 | 0.9759 |
|  | Proposed 1 (MCP) | 0.9797 | 0.9757 | 0.9760 | **0.9808** | 0.9758 | 0.9788 | **0.9671** | 0.9627 | 0.9746 |
| 40% | MH based recovery | 0.9722 | 0.9606 | 0.9721 | 0.9771 | 0.9644 | 0.9237 | 0.9245 | 0.9268 | 0.9527 |
|  | SGSR | 0.9835 | 0.9793 | 0.9784 | 0.9788 | 0.9759 | 0.9799 | 0.9648 | 0.9661 | 0.9758 |
|  | ALSB | 0.9830 | 0.9838 | 0.9830 | 0.9871 | 0.9820 | 0.9730 | 0.9581 | 0.9642 | 0.9768 |
|  | JASR | 0.9803 | 0.9764 | 0.9742 | 0.9808 | 0.9676 | 0.9831 | 0.9739 | 0.9677 | 0.9755 |
|  | GSR-Lp | 0.9854 | 0.9832 | 0.9828 | 0.9866 | 0.9834 | 0.9860 | 0.9767 | 0.9735 | 0.9822 |
|  | GSR-NCR | **0.9879** | **0.9867** | **0.9848** | **0.9886** | **0.9862** | **0.9894** | 0.9794 | **0.9794** | **0.9853** |
|  | Proposed 1 (ETP) | 0.9875 | 0.9860 | 0.9844 | 0.9876 | 0.9848 | 0.9893 | **0.9796** | 0.9791 | 0.9848 |
|  | Proposed 1 (logarithm) | 0.9873 | 0.9857 | 0.9842 | 0.9874 | 0.9845 | 0.9890 | **0.9795** | 0.9785 | 0.9845 |
|  | Proposed 1 (Lp) | 0.9874 | 0.9859 | 0.9843 | 0.9876 | 0.9849 | 0.9892 | **0.9796** | 0.9788 | 0.9847 |
|  | Proposed 1 (MCP) | 0.9874 | 0.9859 | 0.9844 | 0.9877 | 0.9851 | 0.9893 | **0.9795** | 0.9787 | 0.9848 |
| 50% | MH based recovery | 0.9805 | 0.9714 | 0.9795 | 0.9824 | 0.9763 | 0.9465 | 0.9437 | 0.9429 | 0.9654 |
|  | SGSR | 0.9885 | 0.9855 | 0.9844 | 0.9844 | 0.9832 | 0.9872 | 0.9762 | 0.9772 | 0.9833 |
|  | ALSB | 0.9891 | 0.9896 | 0.9883 | 0.9912 | 0.9882 | 0.9854 | 0.9740 | 0.9786 | 0.9856 |
|  | JASR | 0.9902 | 0.9881 | 0.9798 | 0.9865 | 0.9848 | 0.9895 | 0.9818 | 0.9823 | 0.9854 |
|  | GSR-Lp | 0.9904 | 0.9887 | 0.9879 | 0.9908 | 0.9884 | 0.9916 | 0.9839 | 0.9826 | 0.9880 |
|  | GSR-NCR | **0.9919** | **0.9909** | 0.9893 | **0.9921** | **0.9905** | 0.9938 | **0.9871** | 0.9866 | **0.9903** |
|  | Proposed 1 (ETP) | 0.9918 | 0.9907 | 0.9892 | **0.9921** | 0.9903 | **0.9941** | 0.9867 | 0.9865 | 0.9902 |
|  | Proposed 1 (logarithm) | 0.9918 | 0.9908 | **0.9893** | **0.9921** | 0.9903 | **0.9941** | 0.9867 | **0.9866** | 0.9902 |
|  | Proposed 1 (Lp) | 0.9918 | 0.9907 | 0.9892 | **0.9921** | 0.9903 | **0.9941** | 0.9867 | 0.9865 | 0.9902 |
|  | Proposed 1 (MCP) | 0.9918 | 0.9907 | **0.9893** | 0.9919 | 0.9901 | **0.9940** | 0.9867 | 0.9864 | 0.9901 |

To make a visual comparison, we choose three typical images of 'boats', 'leaves' and 'monarch' for 0.1 measurements. All reconstructed images are presented in the **Fig**. 9 to **Fig**. 11. It can be seen that our proposed algorithm can reconstruct image with higher quality, presented in the (h), (i), (j) and (k).

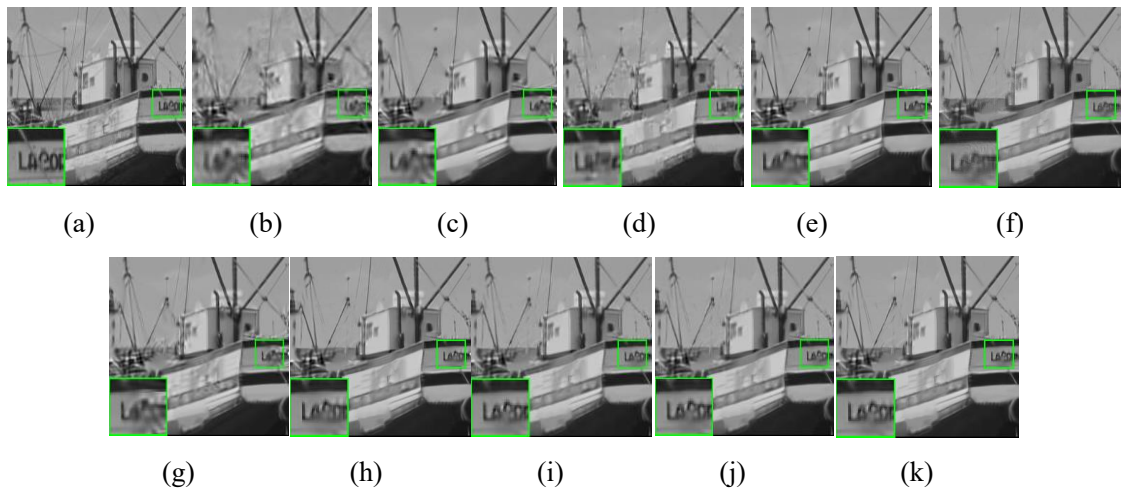

**Fig. 9**. Visual comparison of the original image and seven reconstructed images by SGSR, ALSB, JASR, GSR-Lp, GSR-NCR and our proposed algorithms for 0.1 measurements of boats. (a) Original

image; (b) MH based recovery, 26.09 dB, 0.8481; (c) SGSR, 27.71 dB, 0.8915; (d) ALSB, 27.75 dB, 0.8830; (e) JASR, 28.59 dB, 0.9035; (f) GSR-Lp, 28.37 dB, 0.8983; (g) GSR-NCR, 27.62 dB, 0.8977; (h) Proposed (ETP), **29.03** dB, **0.9076**; (i) Proposed (Logarithm), **29.22** dB, **0.9098**; (j) Proposed (Lp), **29.24** dB, **0.9102**; (k) Proposed (MCP), **29.31** dB, **0.9105**.

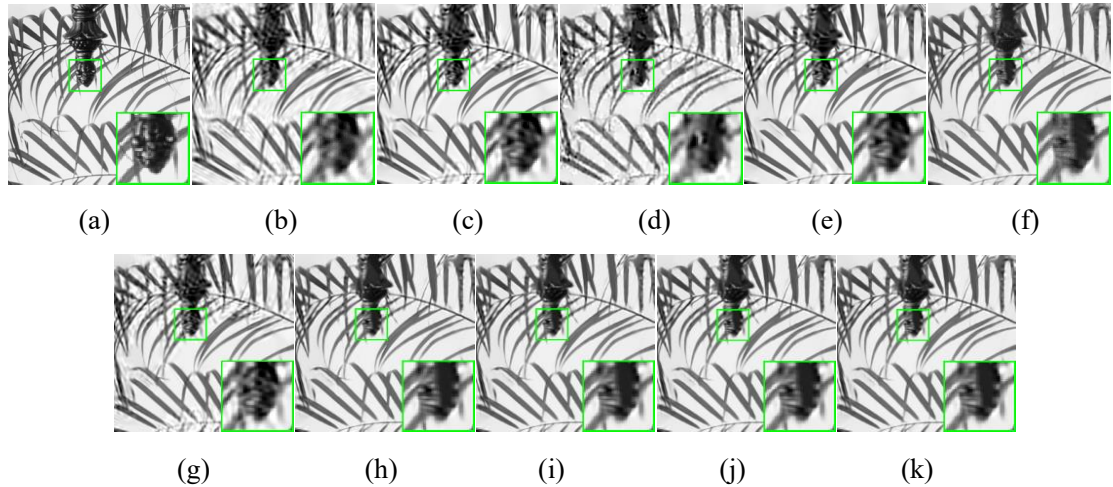

**Fig. 10**. Visual comparison of the original image and seven reconstructed images by SGSR, ALSB, JASR, GSR-Lp, GSR-NCR and our proposed algorithms from 0.1 measurements of leaves. (a) Original Image; (b) MH based recovery, 20.90 dB, 0.7634; (c) SGSR, 22.22, 0.8356; (d) ALSB, 21.37 dB, 0.7934; (e) JASR, 23.62 dB, 0.8799; (f) GSR-Lp, 25.17 dB, 0.9064; (g) GSR-NCR, **21.74** dB, **0.8367**; (h) Proposed (ETP) **25.71** dB, **0.9141**; (i) Proposed (Logarithm) **25.84** dB, **0.9150**; (j) Proposed (Lp), **25.87** dB, **0.9151**; (k) Proposed (MCP), **25.89** dB, **0.9144**.

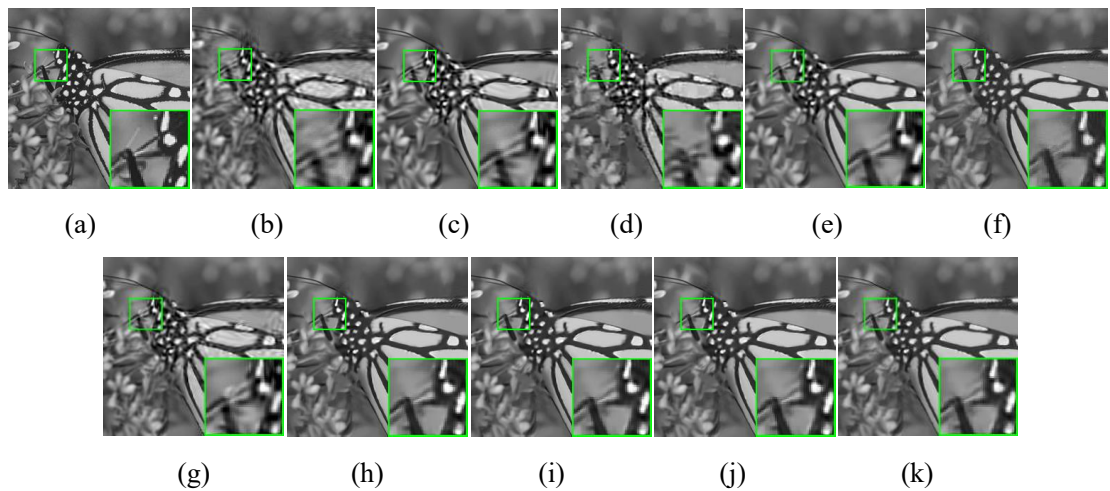

**Fig. 11**. Visual comparison of the original image and five reconstructed images by SGSR, ALSB, JASR, GSR-Lp, GSR-NCR and our proposed algorithms for 0.1 measurements of monarch. (a) Original Image; (b) MH based recovery, 23.20 dB, 0.7912; (c) SGSR, 24.27 dB, 0.8371; (d) ALSB, 24.27 dB, 0.8218; (e) JASR, 25.83 dB, 0.8822; (f) GSR-Lp, 26.61 dB, 0.9003; (g) GSR-NCR, 23.86 dB, 0.8289; (h) Proposed (ETP), **27.29** dB, **0.9066**; (i) Proposed (Logarithm), **27.42** dB, **0.9088**; (j) Proposed (Lp), **27.45** dB, **0.9092**; (k) Proposed (MCP), **27.47** dB, **0.9095**.

**4.1.5 Robustness to Signal-to-Nosie Ratios**

In this subsection, to evaluate the robustness of our proposed framework to the Gaussian noise levels, we conduct reconstruction experiments from CS measurements with three different levels of $\sigma_e = 10$, 20 and 30. We compare the results to four competing methods of MH, SGSR, JASR and GSR-Lp, and examine the performances of our proposed framework for standard CS problems using two nonconvex cases of Logarithm and $L_p$. For a fair comparison, we have carefully adjusted all the parameters for all competing methods. All results are listed in the table **VII**. From the results we can find that our proposed nonconvex method is more robust to Gaussian noise than these convex methods of SGSR and JSAR. For the comparison of GSR-$L_p$ and our proposed method ($L_p$), we can find that our proposed weighting strategy in (39) can avoid the over-shrinking problems effectively, and thus can achieve better performances.

**Table VII** The PSNR/FSIM comparisons of our proposed framework and two competing methods from noisy CS measurements under different levels of Gaussian noise with sampling ratio is 0.2

|  | Method | Barbara | Boats | Elaine | Foreman | House | Leaves | Monarch | Starfish | Average |
|---|---|---|---|---|---|---|---|---|---|---|
| $\sigma_e = 10$ | MH based recovery | 27.76/0.8886 | 27.07/0.8610 | 28.90/0.8892 | 30.33/0.8756 | 29.25/0.8760 | 23.76/0.8139 | 25.23/0.8208 | 24.36/0.8329 | 27.08/0.8573 |
|  | SGSR | 28.95/0.9164 | 28.42/0.8997 | 29.90/0.9099 | 31.80/0.9114 | 30.97/0.9141 | 26.02/0.8862 | 26.66/0.8721 | 25.27/0.8608 | 28.50/0.8963 |
|  | JASR | 29.61/0.9191 | 28.76/0.9033 | 30.38/0.9118 | 33.02/0.9231 | 31.83/0.8984 | 26.72/0.9102 | 27.26/0.8884 | 25.81/0.8740 | 29.17/0.9035 |
|  | GSR-Lp | 29.35/0.9179 | 29.28/0.9095 | 30.75/0.9164 | 33.38/0.9260 | 32.56/0.9152 | 27.21/0.9127 | 28.04/0.9057 | 26.53/0.8902 | 29.63/0.9117 |
|  | Proposed 1 (logarithm) | 29.71/0.9158 | 29.62/0.9092 | 30.96/0.9155 | 33.50/0.9251 | 32.81/0.9138 | 27.79/0.9259 | 28.31/0.9112 | 27.02/0.8951 | 29.97/0.9140 |
|  | Proposed 1 (Lp) | 29.82/0.9189 | 29.68/0.9122 | 31.04/0.9167 | 33.55/0.9219 | 32.65/0.9101 | 27.64/0.9246 | 28.41/0.9152 | 27.11/0.8901 | 29.99/0.9137 |
| $\sigma_e = 20$ | MH based recovery | 23.84/0.8001 | 23.48/0.7653 | 24.67/0.7928 | 25.49/0.7368 | 24.68/0.7588 | 21.48/0.7457 | 22.69/0.7433 | 22.42/0.7841 | 23.59/0.7659 |
|  | SGSR | 25.11/0.8521 | 24.91/0.8273 | 25.81/0.8385 | 27.15/0.8090 | 26.67/0.8327 | 22.90/0.8132 | 23.87/0.8024 | 22.98/0.8097 | 24.93/0.8231 |
|  | JASR | 25.71/0.8601 | 25.86/0.8473 | 26.76/0.8407 | 28.43/0.8564 | 28.95/0.8587 | 24.12/0.8465 | 24.84/0.8201 | 23.78/0.8150 | 26.05/0.8431 |
|  | GSR-Lp | 26.05/0.8643 | 26.31/**0.8601** | 27.50/**0.8733** | 29.93/**0.8829** | 29.40/0.8833 | **24.30**/**0.8780** | 25.34/**0.8636** | 24.13/**0.8440** | 26.62/**0.8687** |
|  | Proposed 1 (logarithm) | 26.31/0.8689 | 26.18/0.8521 | 27.22/0.8591 | 29.58/0.8736 | 28.88/0.8652 | 23.93/0.8630 | 24.69/0.8333 | 24.02/0.8371 | 26.35/0.8565 |
|  | Proposed 1 (Lp) | 26.55/0.8712 | 26.48/0.8575 | 27.56/0.8658 | 30.09/0.8780 | 29.26/**0.8636** | 24.17/0.8676 | 24.91/0.8406 | 24.19/0.8429 | **26.65**/0.8609 |
| $\sigma_e = 30$ | MH based recovery | 21.29/0.7338 | 21.00/0.6937 | 21.70/0.7046 | 22.02/0.6237 | 21.70/0.6650 | 19.51/0.6881 | 20.57/0.6800 | 20.45/0.7279 | 21.03/0.6896 |
|  | SGSR | 22.33/0.7871 | 22.37/0.7621 | 23.01/0.7678 | 24.09/0.7143 | 23.62/0.7499 | 20.50/0.7456 | 21.55/0.7316 | 21.04/0.7593 | 22.31/0.7532 |
|  | JASR | 24.06/0.7987 | 24.37/0.7917 | 25.08/0.8011 | 27.59/0.8343 | 27.11/0.8254 | 22.23/0.8256 | 23.21/0.7912 | 22.04/0.7593 | 24.46/0.8034 |
|  | GSR-Lp | 24.31/0.8248 | 24.78/0.8257 | 25.66/0.8412 | 28.29/0.8568 | 27.65/0.8487 | 22.61/0.8499 | 23.72/0.8322 | 22.77/0.8107 | 24.97/**0.8363** |
|  | Proposed 1 (logarithm) | 24.85/0.8107 | 25.08/0.8039 | 26.09/0.8297 | 29.34/0.8605 | 28.82/0.8439 | 22.70/0.8461 | 23.76/0.8187 | 22.77/0.7917 | 25.43/0.8257 |
|  | Proposed 1 (Lp) | 24.83/0.8120 | 25.24/0.8185 | 26.13/0.8320 | 29.07/0.8600 | 28.63/0.8461 | 23.03/0.8540 | 23.96/0.8251 | 22.99/0.8064 | 25.49/0.8318 |

### 4.2 Robust Image CS under Impulsive noise Environment
#### 4.2.1 Effectiveness of the proposed robust framework

To evaluate our M-estimator based robust CS method, we first generate the noisy measurements by sampling the Fourier transform coefficients randomly [37]. As a typical impulsive noise model, the probability density function (PDF) of Gaussian mixture noise [41] can be given by

$$(1-\xi)N(0,\sigma^2) + \xi(0,\kappa\sigma^2)$$

where $0 \le \xi < 1$ can control the ratio of the outliers, and $\kappa > 1$ indicates the impulsive strength, e.g., $\kappa = 100$. In this paper, we set $\xi = 0.1$ and $\kappa = 100$ to generate the impulsive noise. Because the original image **X** is unknown, here we employ the result of DCT recovery [37] as the initialization for our proposed robust framework. We conduct experiments under five different SNR of 15 dB, 20 dB, 25 dB, 30 dB and 35 dB, which is defined by $SNR =$

$20\log_{10}(\|\mathbf{Ax} - \mathbb{E}(\mathbf{Ax})\|_2/\|\mathbf{n}\|_2)$, where $\mathbf{A}$ denote the measurement matrix, and $\mathbf{x}$ represents the ground truth of sparse signal, the $\mathbf{n}$ denotes the Gaussian mixture noise. To demonstrate the effectiveness of our proposed robust framework, we employ four algorithms as competing methods. The algorithm of DCT is utilized as initialization, L2-based NNM and L2-based GSR-$L_p$ [45] are two GSR based CS method which employ the L2-norm as the fidelity term, the ME-based NNM is a robust CS method which employs the M-estimator to suppress the outliers and utilizes convex nuclear norm as regularizer.

Empirically, we set the block size $\sqrt{\mathcal{B}_s} \times \sqrt{\mathcal{B}_s}$ as $32 \times 32$, the patch size is $6 \times 6$, and the searching window $L \times L$ is $20 \times 20$ for all robust experiments. Table VIII details all the parameters of $\mu$, $\lambda$ and $\eta$ used in our experiments, and $p = 0.5$ for $L_p$-norm. Table IX and Table I summarize the achieved results of PSNR/FSIM, from these results we can see that our proposed robust method can reconstruct image from noisy measurements effectively and can achieve better performances compared with four competing methods.

**Table IX**  The PSNR results of proposed algorithm and competing algorithms (dB)

| SNR | Method | Barbara | Boats | Elaine | Foreman | House | Leaves | Monarch | Starfish | Average |
|---|---|---|---|---|---|---|---|---|---|---|
| 15 dB | DCT based Recovery | 17.74 | 17.60 | 17.49 | 16.47 | 17.45 | 14.31 | 18.39 | 17.60 | 17.13 |
| | L2-based NNM | 19.74 | 19.50 | 19.28 | 18.22 | 19.31 | 16.16 | 20.47 | 19.51 | 19.02 |
| | L2-based GSR-Lp | 24.77 | 24.68 | 25.35 | 26.76 | 26.82 | 20.76 | 24.14 | 23.32 | 24.58 |
| | ME-based NNM | 28.02 | 27.81 | 27.88 | 27.20 | 28.16 | 24.84 | 28.35 | 27.23 | 27.43 |
| | Proposed 2 (logarithm) | **28.38** | **28.09** | **28.73** | **28.36** | **29.26** | **24.75** | **28.58** | **27.09** | **27.84** |
| | Proposed 2 (Lp) | **28.32** | **28.01** | **28.40** | **27.79** | **28.78** | **24.81** | **28.75** | **27.31** | **27.77** |
| | Proposed 2 (MCP) | **28.42** | **28.28** | **28.32** | **27.60** | **28.64** | **25.23** | **28.84** | **27.59** | **27.87** |
| 20 dB | DCT based Recovery | 22.08 | 22.18 | 22.17 | 21.30 | 22.18 | 18.39 | 22.62 | 21.92 | 21.61 |
| | L2-based NNM | 24.81 | 24.55 | 24.53 | 23.67 | 24.69 | 21.18 | 25.31 | 24.30 | 24.13 |
| | L2-based GSR-Lp | 25.49 | 25.19 | 25.33 | 24.76 | 25.56 | 21.91 | 25.38 | 24.75 | 24.80 |
| | ME-based NNM | 31.76 | 31.51 | 31.97 | 31.62 | 32.38 | 28.42 | 31.46 | 29.79 | 31.11 |
| | Proposed 2 (logarithm) | **32.21** | **31.80** | **32.71** | **32.71** | **33.36** | **29.16** | **31.84** | **29.89** | **31.71** |
| | Proposed 2 (Lp) | **32.50** | **32.05** | **32.46** | **31.82** | **32.88** | **29.25** | **32.68** | **31.08** | **31.84** |
| | Proposed 2 (MCP) | **32.21** | **31.89** | **32.53** | **32.37** | **33.08** | **29.19** | **32.07** | **30.10** | **31.68** |
| 25 dB | DCT based Recovery | 25.96 | 26.36 | 26.62 | 26.02 | 26.88 | 21.82 | 26.44 | 25.70 | 25.73 |
| | L2-based NNM | 30.03 | 29.74 | 29.85 | 29.15 | 30.25 | 26.30 | 30.43 | 28.99 | 29.34 |
| | L2-based GSR-Lp | 32.94 | 32.69 | 33.36 | 33.88 | 34.63 | 28.71 | 32.62 | 30.97 | 32.48 |
| | ME-based NNM | 33.92 | 33.67 | 34.27 | 34.27 | 34.73 | 30.66 | 33.58 | 31.95 | 33.38 |
| | Proposed 2 (logarithm) | **35.94** | **35.86** | **36.10** | **36.11** | **36.58** | **32.84** | **35.89** | **33.68** | **35.38** |
| | Proposed 2 (Lp) | **35.71** | **35.48** | **35.89** | **35.83** | **36.39** | **32.14** | **35.83** | **33.81** | **35.14** |
| | Proposed 2 (MCP) | **35.25** | **35.37** | **35.60** | **35.66** | **36.27** | **32.02** | **35.33** | **33.27** | **34.85** |
| 30 dB | DCT based Recovery | 29.15 | 29.93 | 30.43 | 30.50 | 31.04 | 24.37 | 29.39 | 28.59 | 29.18 |
| | L2-based NNM | 34.64 | 34.32 | 35.15 | 35.79 | 36.00 | 31.32 | 34.16 | 32.48 | 34.23 |
| | L2-based GSR-Lp | 35.10 | 34.82 | 35.63 | 36.88 | 36.32 | 33.30 | 34.52 | 32.90 | 34.93 |
| | ME-based NNM | 35.75 | 35.65 | 36.69 | 37.38 | 37.37 | 32.76 | 34.90 | 33.01 | 35.44 |
| | Proposed 2 (logarithm) | **37.74** | **37.92** | **38.44** | **39.73** | **39.37** | **35.27** | **37.24** | **34.05** | **37.47** |
| | Proposed 2 (Lp) | **37.73** | **38.16** | **38.57** | **39.85** | **39.57** | **34.76** | **37.57** | **34.55** | **37.60** |

|  | Proposed 2 (MCP) | **37.28** | **37.69** | **38.08** | **39.21** | **39.30** | **34.09** | **37.09** | **34.27** | **37.13** |
| --- | --- | --- | --- | --- | --- | --- | --- | --- | --- | --- |
| 35 dB | DCT based Recovery | 31.30 | 32.55 | 33.50 | 34.38 | 34.73 | 26.04 | 31.31 | 30.30 | 31.76 |
|  | L2-based NNM | 36.65 | 36.42 | 37.35 | 38.83 | 38.35 | 34.44 | 35.96 | 34.15 | 36.52 |
|  | L2-based GSR-Lp | 38.10 | 38.05 | 38.33 | 39.13 | 39.23 | 35.72 | 38.13 | 36.03 | 37.84 |
|  | ME-based NNM | 36.88 | 36.86 | 37.87 | 39.18 | 38.74 | 33.94 | 36.25 | 34.26 | 36.75 |
|  | Proposed 2 (logarithm) | **38.06** | **38.81** | **39.37** | **41.46** | **40.97** | **35.28** | **38.27** | **35.63** | **38.48** |
|  | Proposed 2 (Lp) | **38.82** | **39.43** | **39.86** | **42.49** | **41.48** | **36.41** | **38.85** | **35.70** | **39.13** |
|  | Proposed 2 (MCP) | **38.21** | **38.65** | **39.39** | **41.86** | **40.70** | **36.44** | **37.86** | **34.83** | **38.49** |

**Table X** The FSIM results of proposed algorithm and competing algorithms

| SNR | Method | *Barbara* | *Boats* | *Elaine* | *Foreman* | *House* | *Leaves* | *Monarch* | *Starfish* | Average |
| --- | --- | --- | --- | --- | --- | --- | --- | --- | --- | --- |
| 15 dB | DCT based Recovery | 0.6500 | 0.6222 | 0.5692 | 0.4418 | 0.5686 | 0.5746 | 0.6201 | 0.6549 | 0.5877 |
|  | L2-based NNM | 0.7093 | 0.6781 | 0.6299 | 0.4996 | 0.6233 | 0.6210 | 0.6842 | 0.7113 | 0.6446 |
|  | L2-based GSR-Lp | 0.8352 | 0.8256 | 0.8465 | 0.8264 | 0.8435 | 0.7877 | 0.8510 | 0.8292 | 0.8306 |
|  | ME-based NNM | 0.8999 | 0.8861 | 0.8737 | 0.8015 | 0.8652 | 0.8331 | 0.8881 | 0.8950 | 0.8678 |
|  | Proposed 2 (Logarithm) | **0.9113** | **0.8965** | **0.8890** | **0.8407** | **0.8864** | **0.8513** | **0.8988** | **0.8951** | **0.8836** |
|  | Proposed 2 (Lp) | **0.9096** | **0.8951** | **0.8828** | **0.8226** | **0.8757** | **0.8465** | **0.8973** | **0.8979** | **0.8784** |
|  | Proposed 2 (MCP) | **0.9076** | **0.8978** | **0.8816** | **0.8153** | **0.8753** | **0.8465** | **0.8990** | **0.9010** | **0.8780** |
| 20 dB | DCT based Recovery | 0.7719 | 0.7515 | 0.7243 | 0.6064 | 0.7058 | 0.6745 | 0.7448 | 0.7741 | 0.7192 |
|  | L2-based NNM | 0.8403 | 0.8148 | 0.7951 | 0.6908 | 0.7804 | 0.7455 | 0.8225 | 0.8364 | 0.7907 |
|  | L2-based GSR-Lp | 0.8542 | 0.8334 | 0.8203 | 0.7314 | 0.8051 | 0.7660 | 0.8375 | 0.8490 | 0.8121 |
|  | ME-based NNM | 0.9471 | 0.9401 | 0.9348 | 0.9015 | 0.9338 | 0.9015 | 0.9376 | 0.9314 | 0.9285 |
|  | Proposed 2 (logarithm) | **0.9525** | **0.9450** | **0.9407** | **0.9210** | **0.9428** | **0.9210** | **0.9431** | **0.9310** | **0.9371** |
|  | Proposed 2 (Lp) | **0.9547** | **0.9474** | **0.9396** | **0.9089** | **0.9399** | **0.9144** | **0.9483** | **0.9435** | **0.9371** |
|  | Proposed 2 (MCP) | **0.9523** | **0.9450** | **0.9397** | **0.9157** | **0.9412** | **0.9187** | **0.9451** | **0.9341** | **0.9365** |
| 25 dB | DCT based Recovery | 0.8625 | 0.8517 | 0.8449 | 0.7623 | 0.8306 | 0.7613 | 0.8422 | 0.8614 | 0.8271 |
|  | L2-based NNM | 0.9298 | 0.9159 | 0.9068 | 0.8512 | 0.9031 | 0.8627 | 0.9221 | 0.9204 | 0.9015 |
|  | L2-based GSR-Lp | 0.9538 | 0.9467 | 0.9429 | 0.9365 | 0.9419 | 0.9200 | 0.9524 | 0.9393 | 0.9417 |
|  | ME-based NNM | 0.9642 | 0.9594 | 0.9565 | 0.9393 | 0.9571 | 0.9310 | 0.9572 | 0.9521 | 0.9521 |
|  | Proposed 2 (logarithm) | **0.9757** | **0.9729** | **0.9677** | **0.9571** | **0.9703** | **0.9551** | **0.9710** | **0.9645** | **0.9668** |
|  | Proposed 2 (Lp) | **0.9748** | **0.9723** | **0.9665** | **0.9552** | **0.9692** | **0.9485** | **0.9704** | **0.9658** | **0.9653** |
|  | Proposed 2 (MCP) | **0.9725** | **0.9701** | **0.9648** | **0.9535** | **0.9680** | **0.9474** | **0.9671** | **0.9623** | **0.9632** |
| 30 dB | DCT based Recovery | 0.9180 | 0.9161 | 0.9135 | 0.8766 | 0.9116 | 0.8213 | 0.8988 | 0.9113 | 0.8959 |
|  | L2-based NNM | 0.9676 | 0.9633 | 0.9601 | 0.9552 | 0.9612 | 0.9480 | 0.9658 | 0.9563 | 0.9597 |
|  | L2-based GSR-Lp | 0.9665 | 0.9596 | 0.9572 | 0.9511 | 0.9424 | 0.9653 | 0.9645 | 0.9520 | 0.9573 |
|  | ME-based NNM | 0.9748 | 0.9727 | 0.9721 | 0.9671 | 0.9743 | 0.9542 | 0.9676 | 0.9605 | 0.9679 |
|  | Proposed 2 (logarithm) | **0.9823** | **0.9815** | **0.9782** | **0.9788** | **0.9821** | **0.9738** | **0.9778** | **0.9664** | **0.9776** |
|  | Proposed 2 (Lp) | **0.9828** | **0.9826** | **0.9790** | **0.9791** | **0.9829** | **0.9710** | **0.9786** | **0.9695** | **0.9782** |
|  | Proposed 2 (MCP) | **0.9815** | **0.9808** | **0.9774** | **0.9767** | **0.9819** | **0.9663** | **0.9769** | **0.9684** | **0.9762** |
| 35 dB | DCT based Recovery | 0.9450 | 0.9485 | 0.9489 | 0.9398 | 0.9542 | 0.8576 | 0.9290 | 0.9343 | 0.9322 |
|  | L2-based NNM | 0.9765 | 0.9735 | 0.9710 | 0.9698 | 0.9688 | 0.9741 | 0.9748 | 0.9662 | 0.9718 |
|  | L2-based GSR-Lp | 0.9826 | 0.9807 | 0.9766 | 0.9743 | 0.9775 | 0.9770 | 0.9810 | 0.9760 | 0.9782 |

| | | | | | | | | | |
|---|---|---|---|---|---|---|---|---|---|
| | ME-based NNM | 0.9799 | 0.9785 | 0.9775 | 0.9773 | 0.9803 | 0.9642 | 0.9746 | 0.9681 | 0.9751 |
| | **Proposed 2 (logarithm)** | **0.9842** | **0.9848** | **0.9824** | **0.9851** | **0.9872** | **0.9730** | **0.9810** | **0.9755** | **0.9817** |
| | **Proposed 2 (Lp)** | **0.9859** | **0.9861** | **0.9835** | **0.9872** | **0.9877** | **0.9791** | **0.9828** | **0.9754** | **0.9835** |
| | **Proposed 2 (MCP)** | **0.9843** | **0.9839** | **0.9820** | **0.9859** | **0.9856** | **0.9795** | **0.9803** | **0.9709** | **0.9816** |

To make a visual comparison, we present the reconstructed images of 'boats', 'elaine' and 'foreman' from noisy measurements corrupted by 15 dB, 20 dB and 25 dB Gaussian mixture noise, respectively. As shown in **Fig**. 12, 13 and 14, we can see that our proposed robust framework can suppress the outliers effectively than other four competing methods.

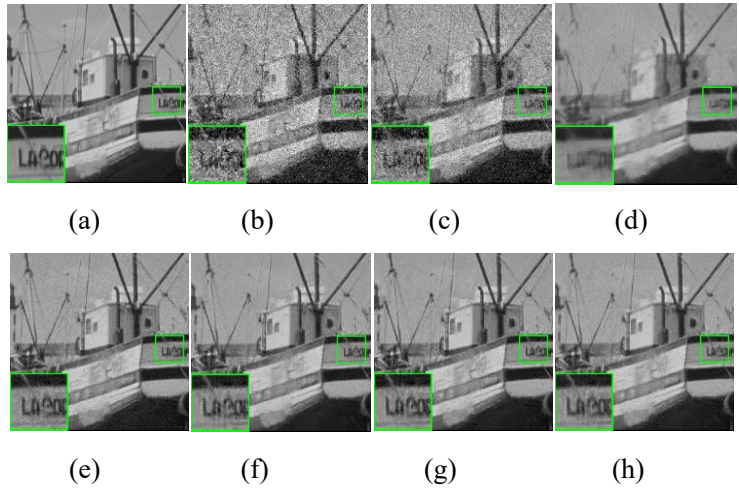

(a)      (b)      (c)      (d)

(e)      (f)      (g)      (h)

**Fig. 12.** Visual comparisons (boats) of the original image and seven reconstructed images from noisy measurement with 15 dB Gaussian mixture noise. (a) Original image. (b) DCT, 17.60 dB, 0.6222; (c) L2-based NNM, 19.50 dB, 0.6781; (d) L2-based GSR-Lp, 24.68 dB, 0.8256; (e) ME-based NNM, 27.81 dB, 0.8861; (f), proposed 2 (Logarithm), 28.09 dB, 0.8965; (g) proposed 2 (Lp), 28.01, 0.8951; (h) proposed 2 (MCP), 28.28dB, 0.8978.

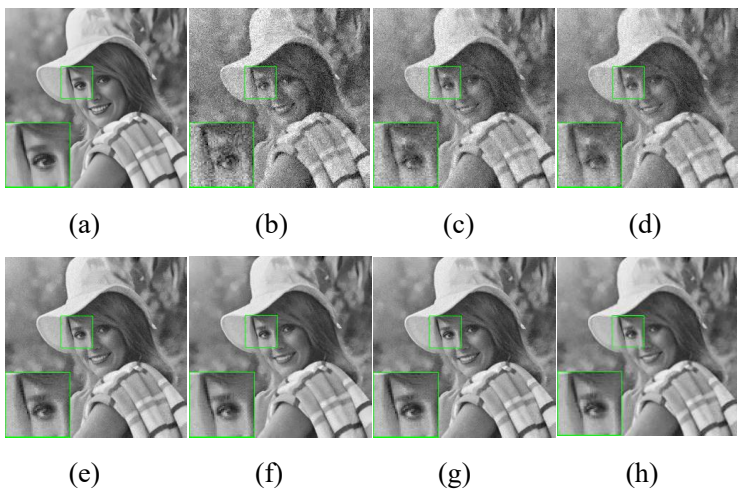

(a)      (b)      (c)      (d)

(e)      (f)      (g)      (h)

**Fig. 13**. Visual comparisons (Elaine) of the original image and seven reconstructed images from noisy measurement with 20 dB Gaussian mixture noise. (a) Original image. (b) DCT, 22.17 dB, 0.7243; (c) L2-based NNM, 24.53 dB, 0.7951; (d) L2-based GSR-Lp, 25.33 dB, 0.8203; (e) ME-

based NNM, 31.97 dB, 0.9348; (f), proposed 2 (Logarithm), 32.71 dB, 0.9407; (g) proposed 2 (Lp), 32.46 dB, 0.9396; (h) proposed 2 (MCP), 32.53 dB, 0.9397.

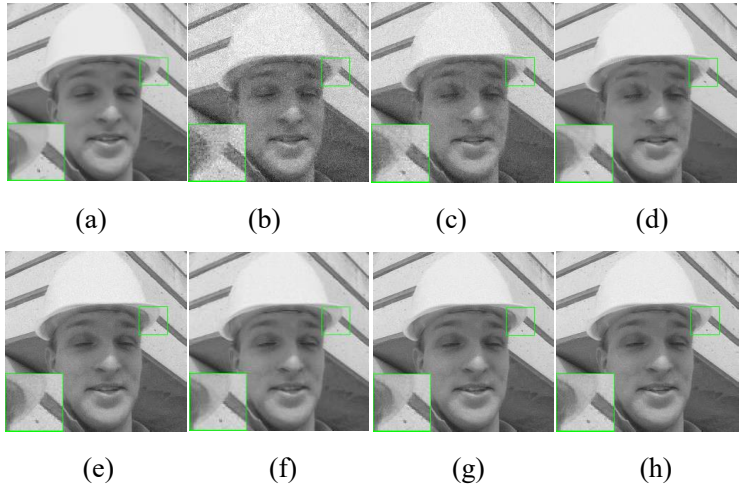

**Fig. 14**. Visual comparisons (Foreman) of the original image and seven reconstructed images from noisy measurement with 25 dB Gaussian mixture noise. (a) Original image. (b) DCT, 26.02 dB, 0.7623; (c) L2-based NNM, 29.15 dB, 0.8512; (d) L2-based GSR-Lp, 33.88 dB, 0.9365; (e) ME-based NNM, 34.27 dB, 0.9393; (f), proposed 2 (Logarithm), 36.11 dB, 0.9571; (g) proposed 2 (Lp), 35.83 dB, 0.9552; (h) proposed 2 (MCP), 35.66 dB, 0.9535.

### 4.2.2 Convergence analysis

Similar to our proposed standard CS algorithm, it is also intractable to demonstrate the convergence of our proposed robust CS algorithm because of the nonconvexity property of regularizer. In this subsection, we will present the convergence property visually by PSNR curves versus the iteration number. **Fig**. 15 (a), (b) and (c) present the PSNRs curves for Logarithm function, Lp function and SCAD function under different sub-sampling rates, from the results we can observe that our proposed algorithm contains good convergence property.

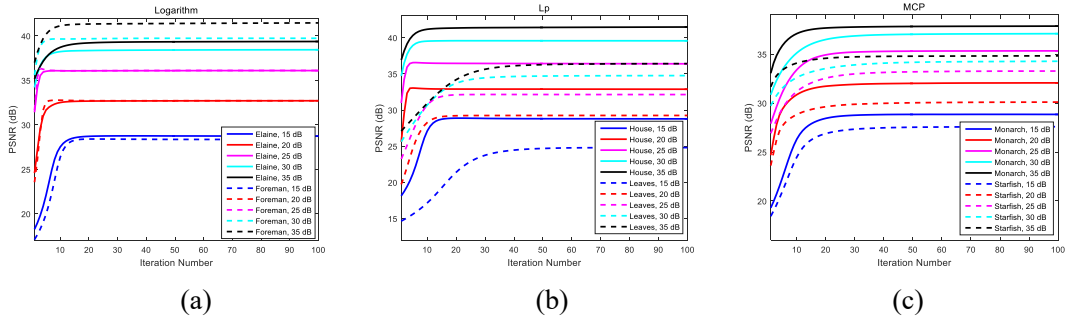

**Fig. 15**. The convergence of the proposed algorithm for ETP function, Logarithm function, MCP function and SCAD function with different SNR of 15 dB, 20 dB, 25 dB, 30 dB and 35dB.

## V. Conclusion

This paper targeted at reconstructing image from compressed sampling data by applying our proposed denoising model, to address the bias problem caused by convex relaxation nuclear norm, some nonconvex surrogates of $L_0$-norm on the singular values are employed to regularize the GSR based low-rank minimization problem. For a better approximation of the rank of group-matrix, an iteratively-weighting strategy $\omega_{k,i}^t = \frac{1}{|\sigma_i(\mathbf{Z}_{G_k}^t)|+\varepsilon}$ is adopted to control the weighting for each singular value. We utilized the popular $L_2$-norm and M-estimator for standard CS and robust CS problems to fit the data, respectively. To solve the resulting optimization problem, we propose a GSR-AIR algorithm. Some conclusions can be achieved from our experimental results:

(1), Compared with the convex NNM, these nonconvex surrogates can improve the reconstruction performance significantly.

(2), The proposed iteratively-weighting strategy $\omega_{k,i}^t = \frac{1}{|\sigma_i(\mathbf{Z}_{G_k}^t)|+\varepsilon}$ is more flexible and effective to control the weighting for each singular value, and hence can achieve good performance.

(3), Compared with the standard CS optimization model, the M-estimator is more effective to suppress the outliers than $L_2$-norm.

## Reference

[1] E.J. Candes, M.B. Wakin, An Introduction to Compressive Sampling, IEEE Signal Process. Mag. 25 (2008) 21–30. doi:10.1109/MSP.2007.914731.

[2] M. Cetin, I. Stojanovic, O. Onhon, K. Varshney, S. Samadi, W.C. Karl, A.S. Willsky, Sparsity-driven synthetic aperture radar imaging: Reconstruction, autofocusing, moving targets, and compressed sensing, IEEE Signal Process. Mag. 31 (2014) 27–40. doi:10.1109/MSP.2014.2312834.

[3] G. Gui, L. Xu, S. Matsushita, Improved adaptive sparse channel estimation using mixed square/fourth error criterion, J. Franklin Inst. 352 (2015) 4579–4594. doi:10.1016/j.jfranklin.2015.07.006.

[4] G. Gui, N. Liu, L. Xu, F. Adachi, Low-complexity large-scale multiple-input multiple-output channel estimation using affine combination of sparse least mean square filters, IET Commun. 9 (2015) 2168–2175. doi:10.1049/iet-com.2014.0979.

[5] G. Gui, H. Huang, Y. Song, H. Sari, Deep Learning for an Effective Nonorthogonal Multiple Access Scheme, IEEE Trans. Veh. Technol. 67 (2018) 8440–8450. doi:10.1109/TVT.2018.2848294.

[6] H. Huang, J. Yang, Y. Song, H. Huang, G. Gui, Deep Learning for Super-Resolution Channel Estimation and DOA Estimation based Massive MIMO System, IEEE Trans. Veh. Technol. 67 (2018) 8549–8560. doi:10.1109/TVT.2018.2851783.

[7] Y. Liu, Globally sparse and locally dense signal recovery for compressed sensing, J. Franklin Inst. 351 (2014) 2711–2727. doi:10.1016/j.jfranklin.2014.01.009.

[8] B. Liu, G. Gui, S. Matsushita, L. Xu, Dimension-reduced direction-of-arrival estimation based on L2,1-norm penalty, IEEE Access. 6 (2018) 44433–44444.

[9] F. Wen, P. Liu, Y. Liu, R.C. Qiu, W. Yu, Robust sparse recovery in impulsive noise via Lp-L1 Optimization, IEEE Trans. Signal Process. 65 (2017) 105–118.

[10] X. Li, Z. Sun, W. Yi, G. Cui, L. Kong, X. Yang, Computationally efficient coherent detection and parameter estimation algorithm for maneuvering target, Signal Processing. 155 (2019) 130–142. doi:10.1016/j.sigpro.2018.09.030.

[11] J. Wen, Z. Zhou, Z. Liu, M.J. Lai, X. Tang, Sharp sufficient conditions for stable recovery of block sparse signals by block orthogonal matching pursuit, Appl. Comput. Harmon. Anal. (2018) 1–27. doi:10.1016/j.acha.2018.02.002.

[12] J. Wen, Z. Zhou, D. Li, X. Tang, A novel sufficient condition for generalized orthogonal matching pursuit, IEEE Commun. Lett. 21 (2017) 805–808. doi:10.1109/LCOMM.2016.2642922.

[13] J. Zhang, Z.L. Yu, Z. Gu, Y. Li, Z. Lin, Multichannel Electrocardiogram Reconstruction in Wireless Body Sensor Networks Through Weighted L1,2 Minimization, IEEE Trans. Instrum. Meas. 67 (2018) 2024–2034. doi:10.1109/TIM.2018.2811438.

[14] J. Cheng, S. Jia, L. Ying, Y. Liu, S. Wang, Y. Zhu, Y. Li, C. Zou, X. Liu, D. Liang, Improved parallel image reconstruction using feature refinement, Magn. Reson. Med. 80 (2018) 211–223. doi:10.1002/mrm.27024.

[15] S. Wang, S. Tan, Y. Gao, Q. Liu, L. Ying, T. Xiao, Y. Liu, X. Liu, H. Zheng, D. Liang, Learning Joint-Sparse Codes for Calibration-Free Parallel MR Imaging, IEEE Trans. Med. Imaging. 37 (2018) 251–261. doi:10.1109/TMI.2017.2746086.

[16] D. Liang, E.V.R. DiBella, R.R. Chen, L. Ying, K-t ISD: Dynamic cardiac MR imaging using compressed sensing with iterative support detection, Magn. Reson. Med. 68 (2012) 41–53. doi:10.1002/mrm.23197.

[17] Z. Zha, X. Liu, X. Huang, H. Shi, Y. Xu, Q. Wang, L. Tang, X. Zhang, Analyzing the group sparsity based on the rank minimization methods, in: IEEE Int. Conf. Multimed. Expo, 2017: pp. 883–888. doi:10.1109/ICME.2017.8019334.

[18] Y. Li, S. Fan, J. Yang, J. Xiong, X. Cheng, G. Gui, S. Hikmet, MUSAI-L1/2: MUltiple Sub-wavelet-dictionaries-based Adaptively-weighted Iterative Half Thresholding Algorithm for Compressive Imaging, IEEE Access. 6 (2018) 16795–16805. doi:10.1109/ACCESS.2018.2799984.

[19] Y. Li, F. Dai, X. Cheng, L. Xu, G. Gui, Multiple-prespecified-dictionary sparse representation for compressive sensing image reconstruction with nonconvex regularization, J. Franklin Inst. 356 (2019) 2353–2371. doi:10.1016/j.jfranklin.2018.12.013.

[20] A. Chambolle, An Algorithm for Total Variation Minimization and Applications, J. Math. Imaging Vis. 20 (2004) 89–97. doi:10.1023/B:JMIV.0000011321.19549.88.

[21] A. Buades, B. Coll, J. Morel, A review of image denoising algorithms , with a new one To cite this version : HAL Id : hal-00271141, Multiscale Model. Simul. 4 (2005) 490–530.

[22] K. Dabov, A. Foi, V. Katkovnik, K. Egiazarian, Image Denoising by Sparse 3-D Transform-Domain Collaborative Filtering, IEEE Trans. Image Process. 16 (2007) 2080–2095.

[23] J. Yang, F. Liu, H. Yue, X. Fu, C. Hou, F. Wu, Textured Image Demoiréing via Signal Decomposition and Guided Filtering, IEEE Trans. Image Process. 26 (2017) 3528–3541.

[24] H. Yue, X. Sun, S. Member, J. Yang, F. Wu, Image Denoising by Exploring External and Internal Correlations, IEEE Trans. Image Process. 24 (2015) 1967–1982. doi:10.1109/TIP.2015.2412373.

[25] W. Dong, L. Zhang, G. Shi, X. Li, Nonlocally centralized sparse representation for image restoration, in: IEEE Trans. Image Process., 2013: pp. 1620–1630. doi:10.1109/TIP.2012.2235847.

[26] J. Zhang, D. Zhao, W. Gao, Group-Based Sparse Representation for Image Restoration, Image Process. IEEE Trans. 23 (2014) 3336–3351. doi:10.1109/TIP.2014.2323127.

[27] Y. Li, Y. Lin, X. Cheng, Z. Xiao, F. Shu, G. Gui, Nonconvex Penalized Regularization for Robust Sparse Recovery in the Presence of SαS Noise, IEEE Access. 6 (2018) 25474–25485. doi:10.1109/ACCESS.2018.2830771.

[28] F. Wen, R. Ying, P. Liu, R.C. Qiu, Robust PCA Using Generalized Nonconvex Regularization, IEEE Trans. Circuits Syst. Video Technol. PP (2019) 1–1. doi:10.1109/tcsvt.2019.2908833.

[29] Y. Li, J. Zhang, S. Fan, J. Yang, J. Xiong, X. Cheng, H. Sari, F. Adachi, G. Gui, Sparse Adaptive Iteratively-Weighted Thresholding Algorithm (SAITA) for Lp-Regularization Using the Multiple Sub-Dictionary Representation, Sensors. 17 (2017) 2920–2936. doi:10.3390/s17122920.

[30] I.E. Frank, J.H. Friedman, A Statistical of Some Chemometrics View Regression Tools, Technometrics. 35 (1993) 109–135. doi:10.2307/1269656.

[31] Z. Xu, H. Zhang, Y. Wang, X. Chang, Y. Liang, $L_{1/2}$ Regularization: A Thersholding Representation Theory and a Fast Solver, IEEE Tansactions Neural Netw. Learn. Syst. 23 (2012) 1013–1027.

[32] J. Fan, R. Li, Variable Selection via Nonconcave Penalized Likelihood and its Oracle Properties, J. Am. Stat. Assoc. 96 (2001) 1348–1360. doi:10.1198/016214501753382273.

[33] J.H. Friedman, Fast sparse regression and classification, Int. J. Forecast. 28 (2012) 722–738. doi:10.1016/j.ijforecast.2012.05.001.

[34] C.H. Zhang, Nearly unbiased variable selection under minimax concave penalty, 2010. doi:10.1214/09-AOS729.

[35] J. Yang, X. Yang, X. Ye, C. Hou, Reconstruction of Structurally-Incomplete Matrices With Reweighted Low-Rank and Sparsity Priors, IEEE Trans. Image Process. 26 (2017) 1158–1172.

[36] Y. Peng, J. Suo, Q. Dai, W. Xu, Reweighted Low-Rank Matrix Recovery and its Application in Image Restoration, IEEE Trans. Cybern. 44 (2014) 2418–2430. doi:10.1109/TCYB.2014.2307854.

[37] W. Dong, G. Shi, X. Li, Y. Ma, F. Huang, Compressive sensing via nonlocal low-rank regularization, IEEE Trans. Image Process. 23 (2014) 3618–3632. doi:10.1109/TIP.2014.2329449.

[38] Q. Wang, X. Zhang, Y. Wu, L. Tang, Z. Zha, Nonconvex Weighted $\ell_p$ Minimization Based Group Sparse Representation Framework for Image Denoising, IEEE Signal Process. Lett. 24 (2017) 1686–1690. doi:10.1109/LSP.2017.2731791.

[39] J. Yang, Y. Zhang, Alternating Direction Algorithms for L1 Problems in Compressive Sensing, SIAM J. Sci. Comput. 33 (2011) 250–278. doi:10.1137/090777761.

[40] Fei Wen, L. Pei, Y. Yang, W. Yu, P. Liu, Efficient and Robust Recovery of Sparse Signal and Image Using Generalized Nonconvex Regularization, IEEE Trans. Comput. Imaging. 3 (2017) 566–579.

[41] D.-S. Pham, S. Venkatesh, Improved image recovery from compressed data contaminated with impulsive noise, IEEE Trans. Image Process. 21 (2012) 397–405. doi:10.1109/TIP.2011.2162418.

[42] Razieh Keshavarzian, A. Aghagolzadeh, Tohid Yousefi Rezaii, LLp Norm Regularization based Group Sparse Representation for Image Compressed Sensing Recovery, Signal Process. Image Commun. 78 (2019) 477–493.

[43] T. Geng, G. Sun, Y. Xu, J. He, Truncated Nuclear Norm Minimization Based Group Sparse Representation, SIAM J. Imaging Sci. 11 (2018) 1878–1897.

[44] R. He, W.S. Zheng, T. Tan, Z. Sun, Half-quadratic-based iterative minimization for robust sparse representation, IEEE Trans. Pattern Anal. Mach. Intell. 36 (2014) 261–275. doi:10.1109/TPAMI.2013.102.

[45] C. Zhao, J. Zhang, S. Ma, W. Gao, Non-convex Lp Nuclear Norm based ADMM framewrok for Compressed Sensing, in: 2016 Data Compression Conf., 2016: pp. 161–170. doi:10.1109/DCC.2016.104.

[46] Z. Zha, X. Zhang, Q. Wang, L. Tang, X. Liu, Group-based sparse representation for image compressive sensing reconstruction with non-convex regularization, Neurocomputing. 296 (2018) 55–63. doi:10.1016/j.neucom.2018.03.027.

[47] C. Gao, N. Wang, A Feasible Nonconvex Relaxation Approach to Feature Selection, in: Proc. Twenty-Fifth AAAI Conf. Artif. Intell., 2011: pp. 356–361.

[48] T. Zhang, Analysis of multi-stage convex relaxation for sparse regularization, J. Mach. Learn. Res. 11 (2010) 1081–1107.

[49] D. Geman, C. Yang, Nonlinear Image Recovery with Half-Quadratic Regularization, IEEE Trans. Image Process. 4 (1995) 932–946. doi:10.1109/83.392335.

[50] J. Trzasko, A. Manduca, Highly undersampled magnetic resonance image reconstruction via homotopic L0-minimization, IEEE Trans. Med. Imaging. 28 (2009) 106–121. doi:10.1109/TMI.2008.927346.

[51] J. Yang, Y. Zhang, Alternating direction algorithms for L1 compressive sensing, SIAM J. Sci. Comput. 33 (2011) 250–278.

[52] M. Nikolova, M.K. Ng, Analysis of half-quadratic minimization methods for signal and image recovery, SIAM J. Sci. Comput. 27 (2005) 937–966. doi:10.1137/030600862.

[53] C. Lu, J. Tang, S. Yan, Z. Lin, Nonconvex nonsmooth low rank minimization via iteratively reweighted nuclear norm, IEEE Trans. Image Process. 25 (2016) 829–839. doi:10.1109/TIP.2015.2511584.

[54] Z. Zha, X. Zhang, Y. Wu, Q. Wang, X. Liu, L. Tang, X. Yuan, Non-convex weighted ℓp nuclear norm based ADMM framework for image restoration, Neurocomputing. 311 (2018) 209–224. doi:10.1016/j.neucom.2018.05.073.

[55] C. Lu, J. Tang, S. Yan, Z. Lin, Generalized nonconvex nonsmooth low-rank minimization, in: Proc. IEEE Comput. Soc. Conf. Comput. Vis. Pattern Recognit., 2014: pp. 4130–4137. doi:10.1109/CVPR.2014.526.

[56] K. Chen, H. Dong, K. Chan, Reduced rank regression via adaptive nuclear norm penalization, Biometrika. 100 (2013) 901–920. doi:10.1093/biomet/ast036.

[57] C. Chen, E.W. Tramel, J.E. Fowler, Compressed-sensing recovery of images and video using multihypothesis predictions, in: 2011 Conf. Rec. Forty Fifth Asilomar Conf. Signals, Syst. Comput., IEEE, 2011: pp. 1193–1198. doi:10.1109/ACSSC.2011.6190204.

[58] L. Zhang, L. Zhang, X. Mou, D. Zhang, FSIM : A Feature Similarity Index for Image Quality Assessment, IEEE Tansactions Image Process. 20 (2011) 2378–2386.

[59] J. Zhang, D. Zhao, F. Jiang, W. Gao, Structural group sparse representation for image Compressive Sensing recovery, in: Data Compression Conf. Proc., IEEE, 2013: pp. 331–340. doi:10.1109/DCC.2013.41.

[60] J. Zhang, C. Zhao, D. Zhao, W. Gao, Image compressive sensing recovery using adaptively learned sparsifying basis via L0 minimization, Signal Processing. 103 (2014) 114–126. doi:10.1016/j.sigpro.2013.09.025.

[61]    N. Eslahi, A. Aghagolzadeh, Compressive Sensing Image Restoration Using Adaptive Curvelet Thresholding and Nonlocal Sparse Regularization, IEEE Tansactions Image Process. 25 (2016) 3126–3140.